\documentclass[letterpaper,journal]{IEEEtran}

\usepackage{amsmath,amsfonts,amssymb}
\usepackage{algorithm}
\usepackage{algpseudocode}
\usepackage{graphicx}
\usepackage{array}
\usepackage{xcolor}
\usepackage[caption=false]{subfig}
\usepackage{textcomp}
\usepackage{url}
\usepackage{cite}
\usepackage{mathrsfs}
\usepackage{booktabs}
\usepackage{afterpage}
\usepackage{enumerate}
\usepackage{stfloats}
\usepackage{flushend}
\usepackage{indentfirst}
\usepackage[colorlinks=true, linkcolor=blue, citecolor=blue, urlcolor=blue, bookmarks=false]{hyperref}
\usepackage{cleveref}

\allowdisplaybreaks
\newcolumntype{C}[1]{>{\centering\arraybackslash}p{#1}}

\renewcommand{\thetable}{\arabic{table}}
\renewcommand{\thefigure}{\arabic{figure}}
\crefformat{figure}{Fig.~#2(#1)#3}

\def\BibTeX{{\rm B\kern-.05em{\sc i\kern-.025em b}\kern-.08em
    T\kern-.1667em\lower.7ex\hbox{E}\kern-.125emX}}

\begin{document}

\renewcommand{\topfraction}{1.0}
\renewcommand{\bottomfraction}{1.0}
\renewcommand{\dbltopfraction}{1.0}
\renewcommand{\textfraction}{0.0}
\renewcommand{\floatpagefraction}{0.9}
\renewcommand{\dblfloatpagefraction}{0.9}
\setcounter{topnumber}{5}
\setcounter{bottomnumber}{5}
\setcounter{totalnumber}{10}
\hyphenation{op-tical net-works semi-conduc-tor IEEE-Xplore}

\title{{\fontsize{24}{28}\selectfont Towards Efficient Low Altitude Sensing:\\ A Dual Heterogeneous Graph Learning Method for \\UAV Task Allocation}}

\author{ Guangyu Lei, Tianhao Liang,~\IEEEmembership{Member,~IEEE}, Bingyan Xie, Tingting Zhang,~\IEEEmembership{Member,~IEEE}\vspace{-20pt}

\thanks{Guangyu Lei, Tianhao Liang, Tingting Zhang are with the School of Information Science and Technology, Harbin Institute of Technology (Shenzhen), Shenzhen 518055, China, and also with the Guangdong Provincial Key Laboratory of Space-Aerial Networking and Intelligent Sensing, Shenzhen, China (email: GuangyuLei@stu.hit.edu.cn, liangth@hit.edu.cn, zhangtt@hit.edu.cn). Bingyan Xie is with the Department of Electronic Engineering, Shanghai Jiao Tong University, Shanghai 200240, China (email: bingyanxie@sjtu.edu.cn).(Corresponding Author: Tianhao Liang, Tingting Zhang)

This manuscript is a preprint and is currently under peer review for journal publication. As it has not yet undergone peer review, certain descriptions, experimental results, analyses, and references may be revised based on reviewer feedback or further validation by the authors. Any identified errors or inaccuracies will be corrected in subsequent versions. Readers are encouraged to refer to the latest version of this manuscript.}}

\markboth{IEEE Transactions on Network Science and Engineering}
{How to Use the IEEEtran \LaTeX \ Templates}

\maketitle

\begin{abstract}
With the development of low altitude intelligent systems, multiple unmanned aerial vehicles (UAVs) can collaboratively execute more complex tasks. Conventional task allocation methods usually regard tasks and UAVs as isolated entities, making it difficult to capture task dependencies and UAV communication relationships. To address the issue, the paper proposes a dual heterogeneous graph learning based UAV task allocation method. A directed task graph is constructed to represent task dependencies and encode task resource requirements, while an undirected UAV graph is built to model communication relationships and encode UAV resource states. The task allocation problem is then formulated as a structural matching problem between the task graph and the UAV communication graph. Additionally, the paper introduces a graph attention network based feature extraction method to learn structural information from both graphs through message passing. A cross attention mechanism is further combined with proximal policy optimization to perform the matching between task nodes and UAV nodes for task allocation optimization. Simulation results demonstrate that the proposed method achieves more completed tasks and shorter task completion time than benchmark methods under different evaluation metrics. Furthermore, an application case for UAV sensing and computing tasks is developed on the AirSim simulation platform. A large language model is employed to convert natural language task requirements into a task graph for autonomous UAV task execution, demonstrating its potential as a backend framework for natural language driven UAV task execution.
\end{abstract}


\section{Introduction}
\subsection{Background}
\IEEEPARstart{W}{ith} the rapid development of the low altitude economy and intelligent sensing technologies, unmanned aerial vehicles (UAVs) have gradually evolved from simple flying platforms into autonomous task execution units \cite{Qingquan, Ding, TianhaoIOTJ1, TianhaoTWC, zhangxj, Ma, TianhaoVTC, TianhaoVTC2, TianhaoWCNC}. Compared with a single UAV, a UAV swarm can accomplish more complex tasks and provide higher operational reliability in challenging environments \cite{GNSS, TianhaoNET2, TianhaoICC}. Consequently, exploiting the collaborative capabilities of UAV swarms has become an important topic in low altitude intelligent systems \cite{LeiWCL, TianhaoNET}.

As task scales continue to increase, the organization of low altitude sensing tasks has become increasingly complex \cite{TianhaoWCL, TangTWC2025}. In practical applications, tasks are rarely composed of isolated targets. Instead, they are usually formed by multiple subtasks connected through logical relationships \cite{FerreiraAccess2024, ZhengAccess2024}. A regional sensing task often requires information collection before triggering further analysis in other areas according to the obtained results \cite{IOTM, ZhengAccess2024, WangOJCOMS2025}. Treating tasks as independent entities and directly assigning them to UAVs cannot adequately represent execution order or data dependencies among tasks \cite{FerreiraAccess2024, WangOJCOMS2025}. Such a strategy also makes it difficult to guarantee the successful completion of complex task workflows. Therefore, a more structured task representation is required to enable UAV systems to understand the internal logical structure of complex tasks.

Building upon this requirement, collaborative task execution among multiple UAVs must also address resource selection and coordination decisions \cite{GuoTWC2024, HaoTMC2024, Zeng, YuqiGC}. UAVs often possess heterogeneous resources and capabilities, while different tasks exhibit diverse requirements in terms of timing constraints, data volume, and execution difficulty \cite{TianhaoJSAS, HaoTMC2024, ZhengAccess2024}. Task allocation based solely on a single capability metric is insufficient for achieving globally effective decisions. In sensing and computing tasks, UAVs are required not only to reach designated areas and collect data but also to distribute the collected data to other UAVs for cooperative processing \cite{WeiJCIN2024, GuoTWC2024, WCM}. Consequently, task execution should jointly consider task structures, communication conditions, and resource availability, allowing UAV swarms to achieve more efficient coordination under complex task constraints \cite{GuoTWC2024, HaoTMC2024, WeiJCIN2024}.

At the same time, the continuous growth of swarm size and task complexity has reduced the effectiveness of conventional approaches that rely on predefined task templates \cite{C6}. In practical applications, task requirements are often specified in natural language by human operators \cite{CladeraTFR2025, CoTTL2024}. Therefore, a task representation is needed that can align with human intent while capturing complex logical relationships among tasks \cite{yuqistl, CoTTL2024, TR2MTL2024}. Such a representation should accurately describe temporal constraints and dependency relationships between tasks. It should also provide a verifiable and reasoning oriented logical interface for natural language instructions \cite{CoTTL2024, TR2MTL2024}.

Based on the above considerations, there is an urgent need to develop a complex task representation framework that can jointly characterize global task information and UAV network states. Such a framework should systematically model logical dependencies among tasks while establishing a closed loop connection with natural language instructions. On the basis of global feature representations for both task nodes and execution nodes, decision making can be performed more effectively. Furthermore, the feasibility and execution performance of the resulting decisions should be continuously evaluated throughout the task execution process.

\begin{figure*}[t]
\centering
\includegraphics[width=0.9\textwidth]{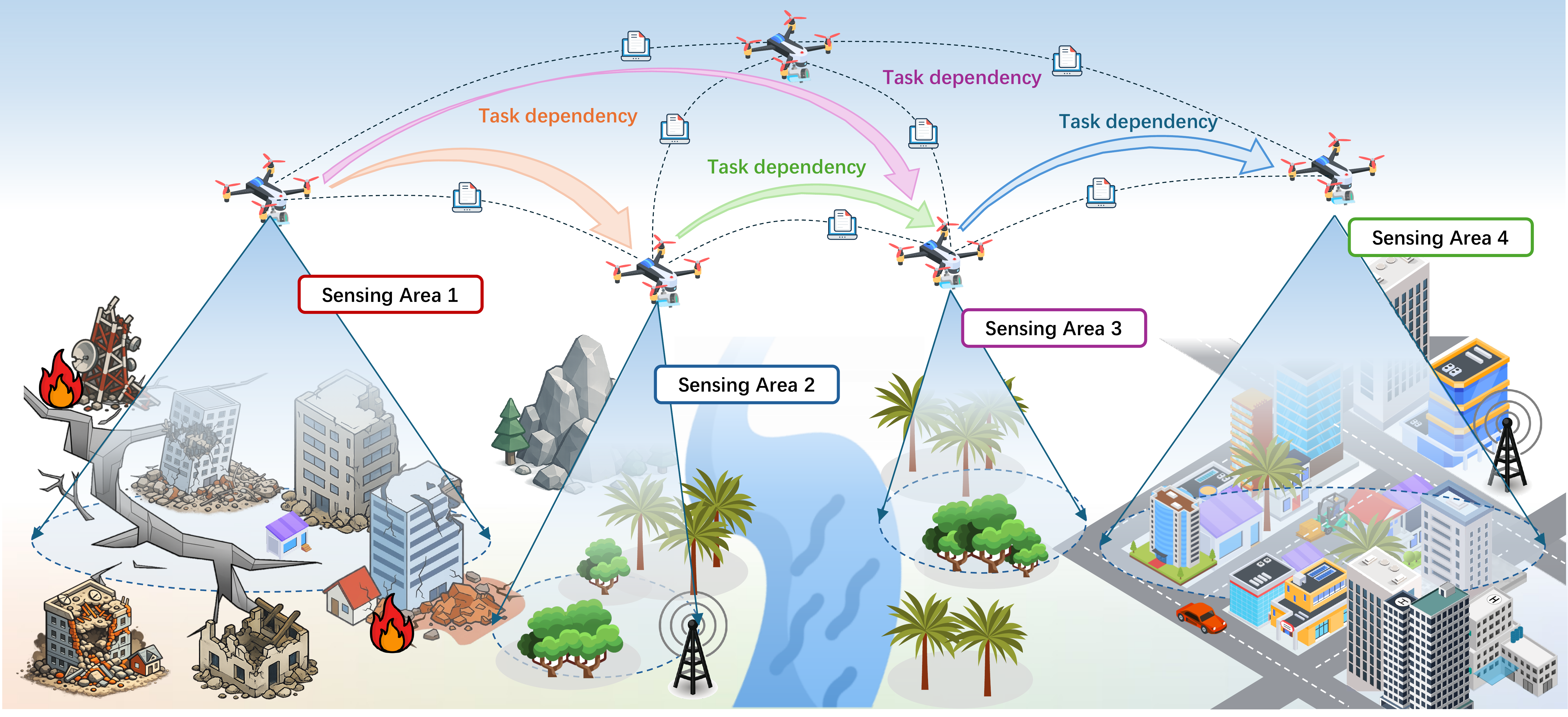}
\vspace{-8pt}
\caption{Low altitude sensing and computing scenario with multiple UAVs, task dependency relations and sensing areas.}
\label{figScenario}
\vspace{-8pt}
\end{figure*}
\vspace{-10pt}
\subsection{Related works}
\subsubsection{Task Allocation Modeling and Representation}
Task allocation among multiple UAVs is commonly formulated as a matching problem between UAVs and tasks. In these formulations, UAVs are modeled as execution entities constrained by position, energy, and capability, while tasks are modeled as independent targets with reward, cost, and deadline attributes. The allocation process is described in \cite{A1} as the association between task requirements and UAV capabilities, with the objective of improving task utility or reducing execution cost \cite{HaiTSMC2025}. Dynamic environments are further considered in \cite{A2}, where task appearance, environmental variation, and UAV state changes require repeated matching decisions \cite{WuTVT2024}. Search and rescue scenarios are reviewed in \cite{A3}, where task urgency and spatial distribution affect allocation results. A collaborative allocation method for search and rescue is proposed in \cite{A14}, where execution reward and path cost are jointly considered.

In most existing works, tasks are still modeled as independent targets or task sets, while optimization algorithms, reinforcement learning, or dynamic reassignment strategies are employed to determine the mapping between UAVs and tasks. Hybrid optimization and learning based heuristics have been adopted to improve allocation efficiency and solution quality \cite{A11,A12,A19}. Deep reinforcement learning has been introduced to learn adaptive allocation policies under dynamic environments \cite{A4,A15}. Dynamic reassignment and distributed negotiation mechanisms have also been developed to improve robustness against task variations and communication constraints \cite{A5,A9,A13}. Nevertheless, tasks are generally represented as independently selectable, visitable, or reassignable entities, and task evolution is mainly characterized by changes in task sets, task attributes, or assignment relationships. The precedence dependencies, triggering relationships, and data flows among subtasks in complex sensing missions are rarely modeled explicitly.

More recently, attempts have been made to incorporate richer task relationships into UAV task allocation. Motion feasibility has been integrated through Dubins path constraints for heterogeneous disaster inspection missions \cite{A6}. Dynamic task allocation has been extended to uncertain environments where UAV decisions are continuously adjusted according to changing task states \cite{A10}. Sequential allocation strategies and temporal task chains have also been investigated to characterize execution order among tasks \cite{A16,A17}. Furthermore, graph theory has been introduced to assist allocation optimization in a two stage framework \cite{A18}. These works indicate that increasing attention has been paid to task ordering and task association during mission execution. However, the introduced relationships are still treated as auxiliary constraints for scheduling or optimization, rather than as an intrinsic representation of the task itself. A unified directed acyclic task graph capable of representing precedence dependencies, trigger relationships, and data transfer relationships remains absent.

\subsubsection{Graph Structure}
Graph structures have been introduced to characterize interactions in multi UAV systems \cite{MaTIV2025}. Heterogeneous task allocation for multiple agents is modeled as a graph in \cite{B1}, where graph neural networks and ant colony optimization are combined to extract relational features among agents. In UAV aided smart agriculture, heterogeneous graph neural networks are combined with reinforcement learning in \cite{B2} for task offloading and resource decisions. Attention augmented inverse reinforcement learning with graph convolutions is applied in \cite{B3} for multi agent task allocation, allowing cooperative policies to be learned from agent relations. The MAGNNET framework is proposed in \cite{B4}, where graph neural networks are used to support allocation under communication constraints.

%These methods show that graph neural networks can represent adjacency relations among UAVs or agents, but graph structures are mainly used for execution entities, while tasks are retained as allocation objects or optimization targets \cite{MaJIOT2024}.

Communication topologies have also been explicitly modeled as graphs in UAV networks. Graph neural networks are used in \cite{B5} for spectrum resource optimization, where nodes and edges represent resource and interference relations. Graph neural networks are combined with multi agent reinforcement learning in \cite{B6} for UAV tactical decision making, enabling UAV interactions to be encoded during policy learning. Deep graph reinforcement learning is applied in \cite{B7} to UAV enabled secure communication with multiple users, where structural relations among communication links and user services are modeled. Multi hop differential topology algorithms are proposed in \cite{B8,B9} to restore connectivity in resilient UAV swarm networks. These works improve the representation of communication and cooperation relations among UAVs, but task side structures are not modeled at the same level. When dependency relations exist among tasks, the communication graph alone is insufficient to evaluate whether the current topology can support task collaboration.

Task dependency graphs have been used in UAV edge computing and task offloading. Dependency aware task offloading in collaborative mobile edge computing assisted by multiple UAVs is investigated in \cite{B10}, where computation tasks are divided into subtasks and dependency relations constrain execution order. Collaborative offloading and resource allocation with task dependencies are considered in \cite{B11} for UAV enabled mobile edge computing, allowing dependencies to affect offloading and resource decisions. In device edge cloud computing, improved actor critic deep reinforcement learning is used in \cite{B12} for dependency aware computation offloading, where a task graph constrains the computation workflow. A computing power pool is introduced in \cite{B13} to support dependent tasks in UAV enabled mobile edge computing. These methods demonstrate that directed acyclic graphs can represent precedence relations in computation tasks \cite{WuJIOT2025}. However, task dependency graphs are mainly treated as offloading constraints, while structural consistency between the communication graph and the task graph is not emphasized.

%Multiple UAV edge computing has further considered trajectory, offloading, and resource allocation jointly \cite{QinTNSE2024}. An air ground collaborative mobile edge computing system assisted by multiple UAVs is proposed in \cite{B14}, where UAVs provide aerial computation support during offloading. For integrated sensing, communication, and edge computing, UAV trajectories and resource allocation are jointly optimized in \cite{B15}, so that task execution is affected by sensing demands and communication resources. Task offloading and trajectory optimization assisted by multiple UAVs are investigated in \cite{B16}, where offloading decisions are designed together with UAV motion. Deep reinforcement learning is used in \cite{B17} for task offloading in edge computing with multiple UAV assets, with emphasis on computation resource utilization. These methods include computation and communication in task execution, but task structures and communication structures are not unified within a dual graph matching framework. The data interaction required by tasks and the communication support provided by UAVs are still handled separately.

Graph structures have also been extended to post disaster task planning, digital twin network management, and multi tier UAV networks. Post disaster planning for multiple UAVs is addressed in \cite{B18} with graph neural networks, allowing task states in disaster scenarios to be represented through graphs. Task driven resource management in multi UAV networks is investigated in \cite{B19} from a digital twin perspective, where virtual network states support physical system decisions.
% Blockchain and digital twin techniques are integrated in \cite{B20} for multi tier UAV networks to support resource coordination and network management across layers. 
These works demonstrate that graphs can provide a unified representation for both tasks and executing agents, where edges capture task dependencies as well as collaboration and communication relationships among agents.

\subsubsection{Interface Between Natural Language Instructions and UAV Task Allocation}
The emergence of large language models (LLMs) has narrowed the semantic gap between natural language instructions and UAV task execution \cite{CuiICCA2024}. A method combining LLMs with safe motion planning is presented in \cite{C1}, where language descriptions are converted into formation and motion plans for robot swarms. This idea is extended to drone swarm choreography in \cite{C4}, where language instructions are transformed into UAV trajectories under safety constraints. FlockGPT is introduced in \cite{C3} for UAV flocking through linguistic orchestration, with outputs focused on group motion and formation control. A prompt driven planning method based on LLMs is proposed in \cite{C2}, allowing multi drone systems to generate initial task plans from natural language. These methods show that natural language can provide high level inputs for UAV systems, but their outputs are usually waypoints, action sequences, or formation plans rather than structured task graphs for task allocation \cite{PueyoIROS2024}. This limits the interpretability of task representations and the optimizability of LLM outputs for multi agent collaborative decision making.

LLMs and large small artificial intelligence frameworks have also been applied to control and planning in robotic systems \cite{KannanIROS2024, zhou2026lsailargesmallai}. Natural language is incorporated into multi robot behavior generation in \cite{C7} through an LLM based control system. Distributed on site knowledge is integrated into multi robot task planning in \cite{C17} for multi object retrieval tasks. Automatic construction of mixed integer linear programming models is presented in \cite{C18}, where natural language descriptions are transformed into optimization formulations for task allocation and scheduling. The role of LLMs in multi robot systems is reviewed in \cite{C16}, where task decomposition, plan generation, and human robot interaction are identified as key functions. These methods connect natural language with task planning, but the outputs still need to be organized as task graphs with dependency relations before being combined with UAV communication topology and computation resources \cite{ObataLRA2025}.

Vision language navigation and aerial agents further extend natural language enabled UAV intelligence. Embodied aerial agents for city scale vision language navigation are presented in \cite{C10}, where language and visual information are used for UAV navigation. Conditional transformers and prompt based text rephrasing are designed in \cite{C11} for language guided quadcopter navigation. NavAgent is proposed in \cite{C12} through multi scale urban street view fusion. End to end vision language navigation is explored in \cite{C13}, where perception, language understanding, and action generation are integrated. CityNavAgent is introduced in \cite{C14} for aerial navigation through hierarchical semantic planning in urban environments.
% These methods mainly focus on navigation by a single UAV or a small number of UAVs, and their outputs are usually paths or actions rather than subtask dependencies or data transfer relations.

Multimodal large models have also been introduced into UAV and aerial agent systems. The integration of LLMs with agentic low altitude mobility is reviewed in \cite{C5}, where LLMs are considered useful for high level task understanding and decision making. The potential of multimodal LLMs for UAV swarms is discussed in \cite{C6}, where perception understanding, collaborative planning, and autonomous decision making are enhanced in a unified model. AerialVLA combines vision, language, and action modeling for aerial agents in \cite{C15}. AutoFly builds a vision language action model for UAV navigation in \cite{C19}, enabling navigation actions to be generated from multimodal inputs. A survey on vision language navigation for aerial robots further shows that language interfaces are becoming an important direction for UAV intelligence \cite{C20}. 
%These methods mainly address mappings among language, vision, and actions.

A clear gap remains between natural language driven UAV methods and conventional UAV task allocation. The former can generate waypoints, action scripts, or high level plans from language \cite{C1}, \cite{C4}, while the latter usually assumes that a task set has already been given and directly solves the allocation relation between UAVs and tasks \cite{A1}, \cite{A2}. Human task intent can be preliminarily understood by LLMs, but the generated information has not been naturally organized into structured task graphs required by task allocation among multiple UAVs. However, most of these works focus on single UAV scenarios and provide limited solutions for multi UAV coordination. Directly using LLMs to generate executable instructions for multiple UAVs can be overly complex, while resource relationships cannot be explicitly represented or optimized, thereby weakening both interpretability and optimizability.

\subsection{Motivation and Main Contributions}

Three main issues remain in the above existing works, which are elaborated as follows.

\begin{enumerate}

\item Existing works lack a unified representation of both execution agents and task resource requirements. Although the capabilities of UAVs and the attributes of tasks are usually considered during task allocation, the logical dependencies among subtasks are rarely modeled explicitly.

\item Existing works based on graph structures have formulated task allocation as a matching problem to a certain extent. However, the communication relationships among UAVs and the allocation relationships between tasks and UAVs are usually modeled separately. The consistency between the communication graph and the task graph is therefore not considered.

\item The emergence of LLMs has made it possible to transform natural language instructions into collaborative behaviors of UAV swarms. However, the existing UAV swarm task allocation works still lack a natural interface for human language instructions. Consequently, natural language descriptions cannot be seamlessly converted into structured task representations that are suitable for subsequent task allocation and cooperative computation.

\end{enumerate}

To address the issues, the main contributions can be summarized as follows.
\begin{enumerate}
\item The paper proposes a dual heterogeneous graph based representation for tasks and UAVs. A directed task graph models task dependencies and task resource requirements, while an undirected UAV graph captures communication relationships and UAV resource states. Additionally, the paper transforms the UAV task allocation problem into a structural matching problem between the task graph and the UAV graph.

\item The Graph Attention Network (GAT) based feature extraction method is introduced to learn structural information from the task graph and the UAV graph. Proximal Policy Optimization (PPO) is combined with the cross attention mechanism to match task nodes with UAV nodes, optimizing task allocation for more completed tasks and shorter task completion time.

\item The simulation results show that, compared with benchmark methods, the proposed method achieves better performance under different task numbers and task complexity levels, demonstrating its effectiveness across different evaluation metrics.

\item An application case for UAV sensing and computing tasks on the AirSim simulation platform is introduced. An LLM is employed to map natural language task requirements into a task graph, enabling the UAV swarm to execute sensing and computing tasks. The inference results further verify the feasibility of the proposed method and demonstrate its potential as a natural language interface for UAV task execution.
\end{enumerate}

\section{System Model}
\subsection{Heterogeneous Graph Framework}

We consider a low altitude sensing and computing system composed of multiple UAVs and multiple sensing task nodes. The UAV set is denoted as $\mathcal{U}=\{1,\ldots,N\}$ and the task node set is denoted as $\mathcal{M}=\{1,\ldots,M\}$. To jointly describe task execution dependencies and UAV communication cooperation, the system state is modeled as a heterogeneous graph $\mathcal{G}$. The heterogeneous graph consists of the UAV node set $\mathcal{U}$, the task node set $\mathcal{M}$, the UAV communication edge set $\mathcal{E}_{\mathrm{UU}}$ and the task dependency edge set $\mathcal{E}_{\mathrm{TT}}$. It is given by
\begin{equation}
\mathcal{G}=\left(\mathcal{U},\mathcal{M},\mathcal{E}_{\mathrm{UU}},\mathcal{E}_{\mathrm{TT}}\right).
\label{eqHeterogeneousGraph}
\end{equation}
Here, $\mathcal{E}_{\mathrm{UU}}$ denotes the candidate communication links among UAVs, and $\mathcal{E}_{\mathrm{TT}}$ denotes the directed edges between tasks. The task structure and the UAV network structure are represented within a unified graph model, which provides a common state input for task allocation, topology generation and computation splitting decisions.

\begin{figure*}[!t]
\centering
\subfloat[Complexity $0.05$]{\includegraphics[width=0.24\textwidth]{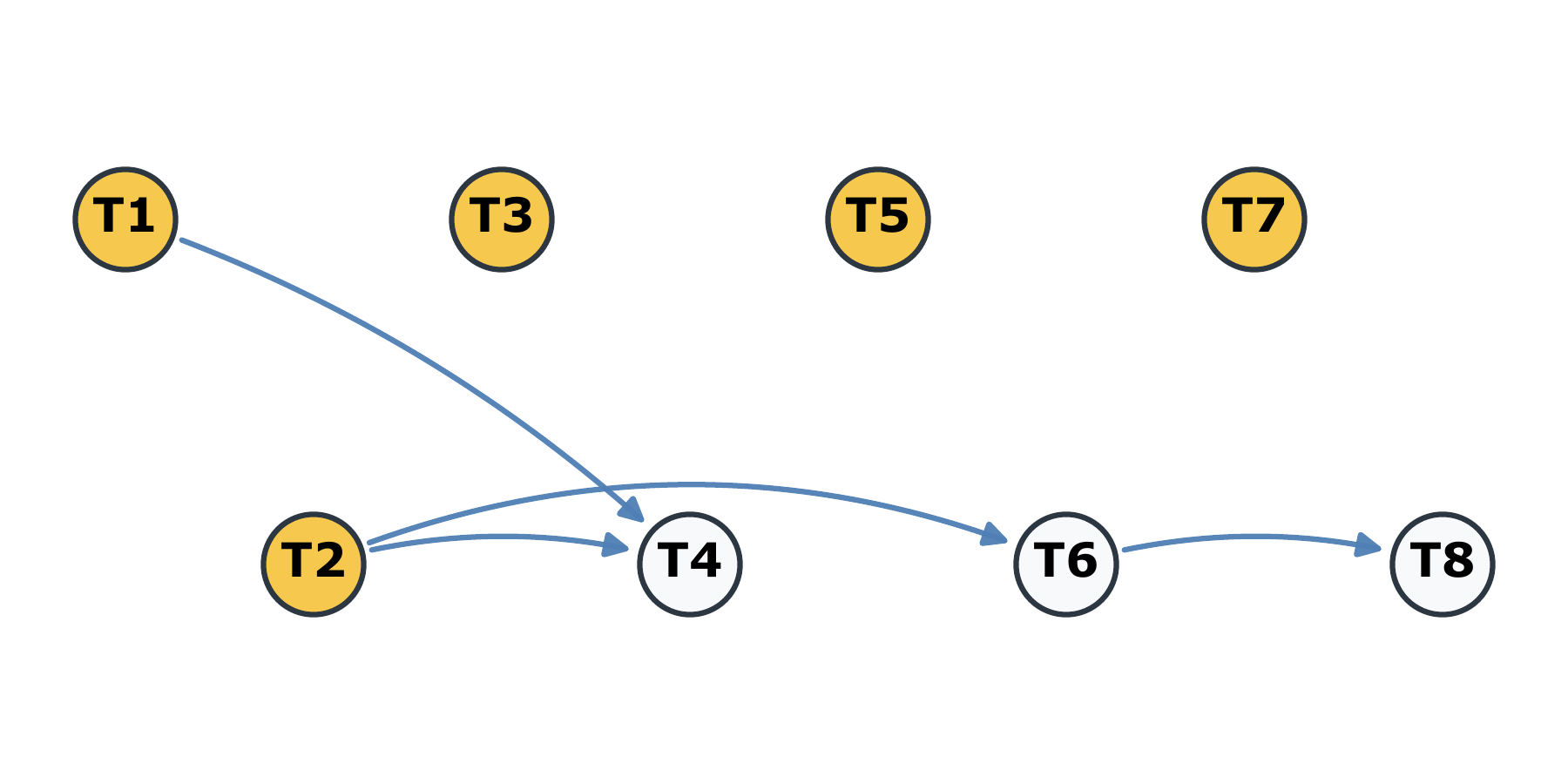}
\label{figTaskGraphC05}}
\subfloat[Complexity $0.15$]{\includegraphics[width=0.24\textwidth]{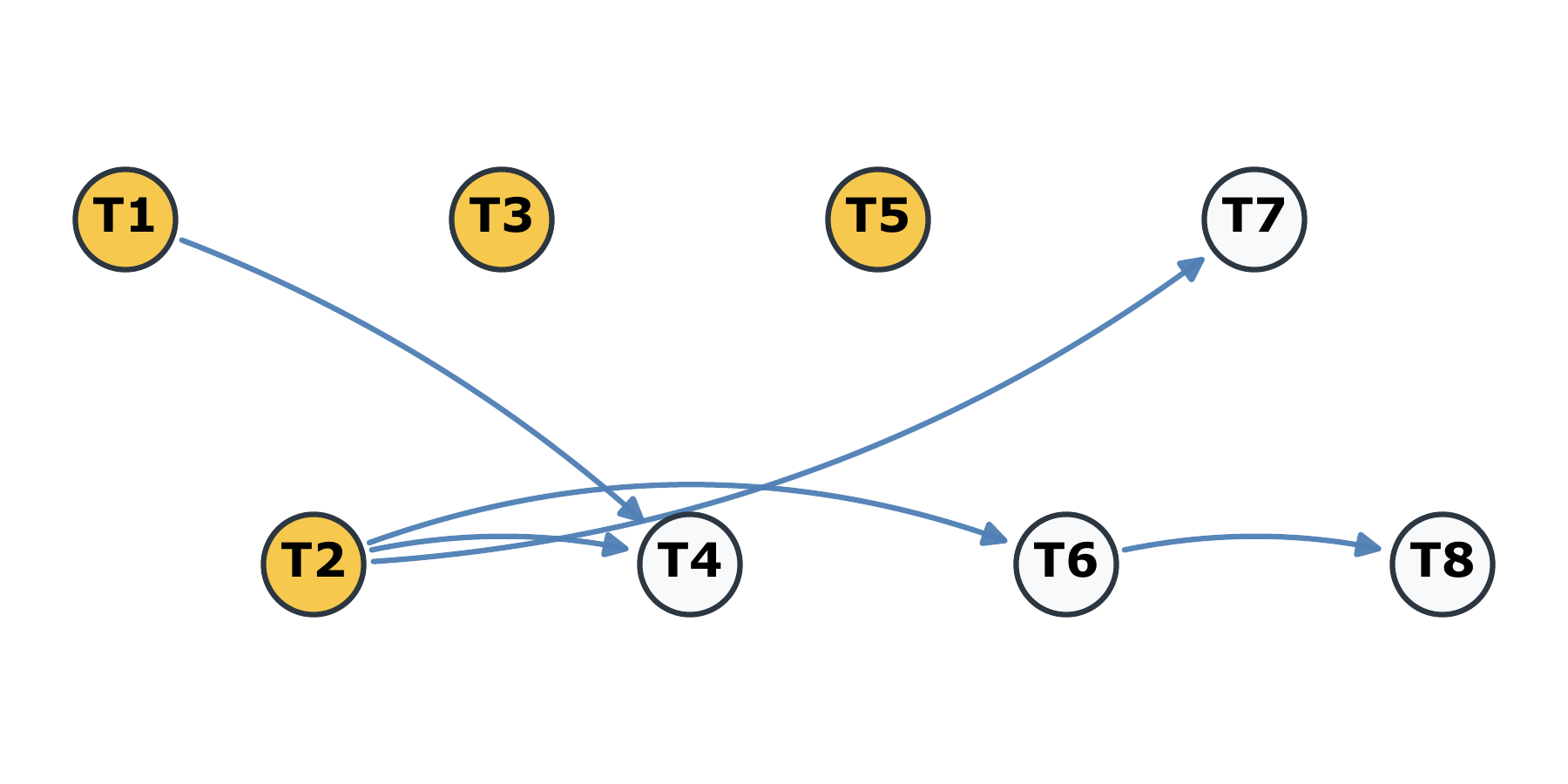}
\label{figTaskGraphC15}}
\subfloat[Complexity $0.25$]{\includegraphics[width=0.24\textwidth]{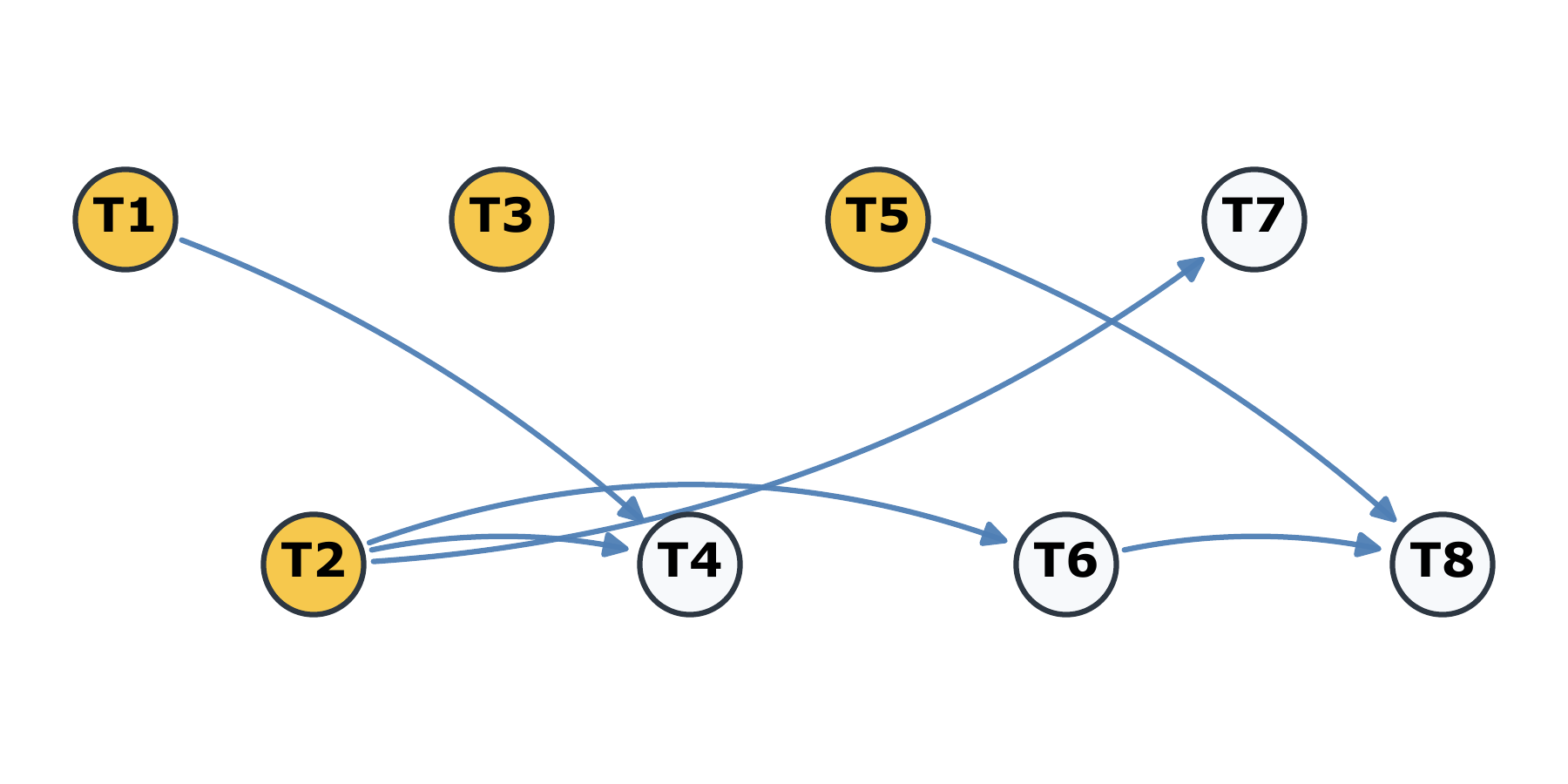}
\label{figTaskGraphC25}}
\subfloat[Complexity $0.35$]{\includegraphics[width=0.24\textwidth]{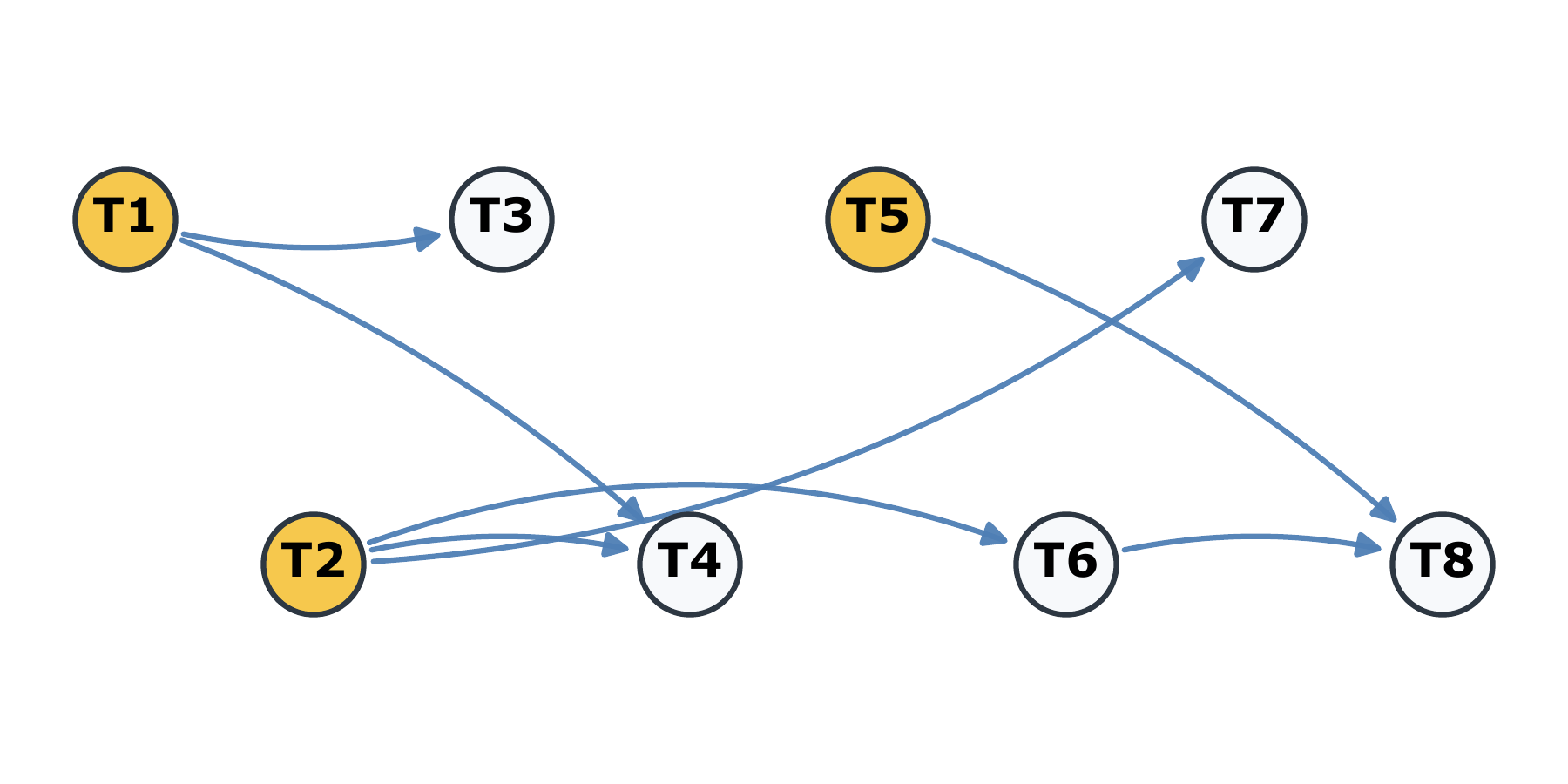}
\label{figTaskGraphC35}}\\ \vspace{-8pt}
\subfloat[Complexity $0.45$]{\includegraphics[width=0.24\textwidth]{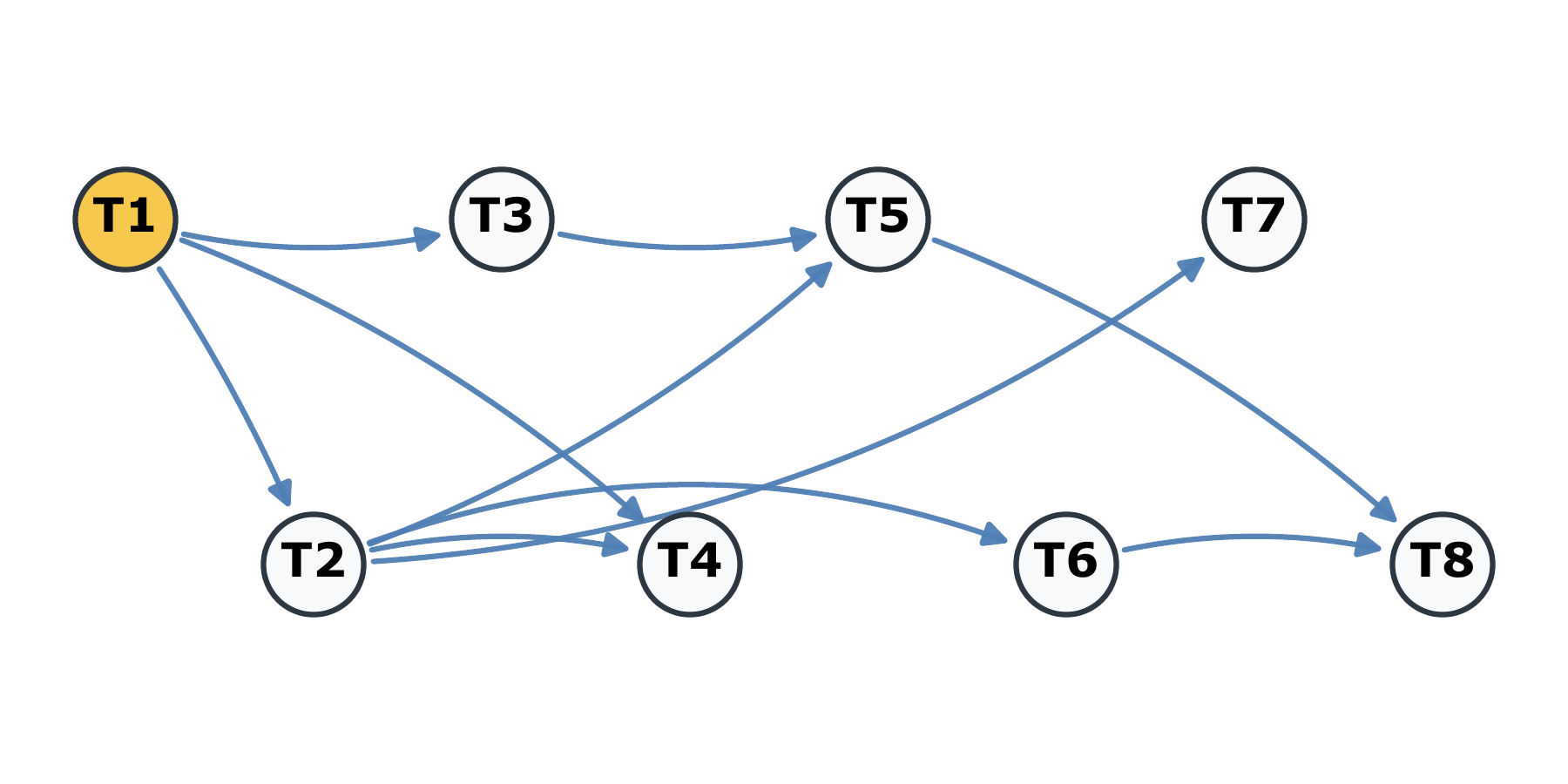}
\label{figTaskGraphC45}}
\subfloat[Complexity $0.55$]{\includegraphics[width=0.24\textwidth]{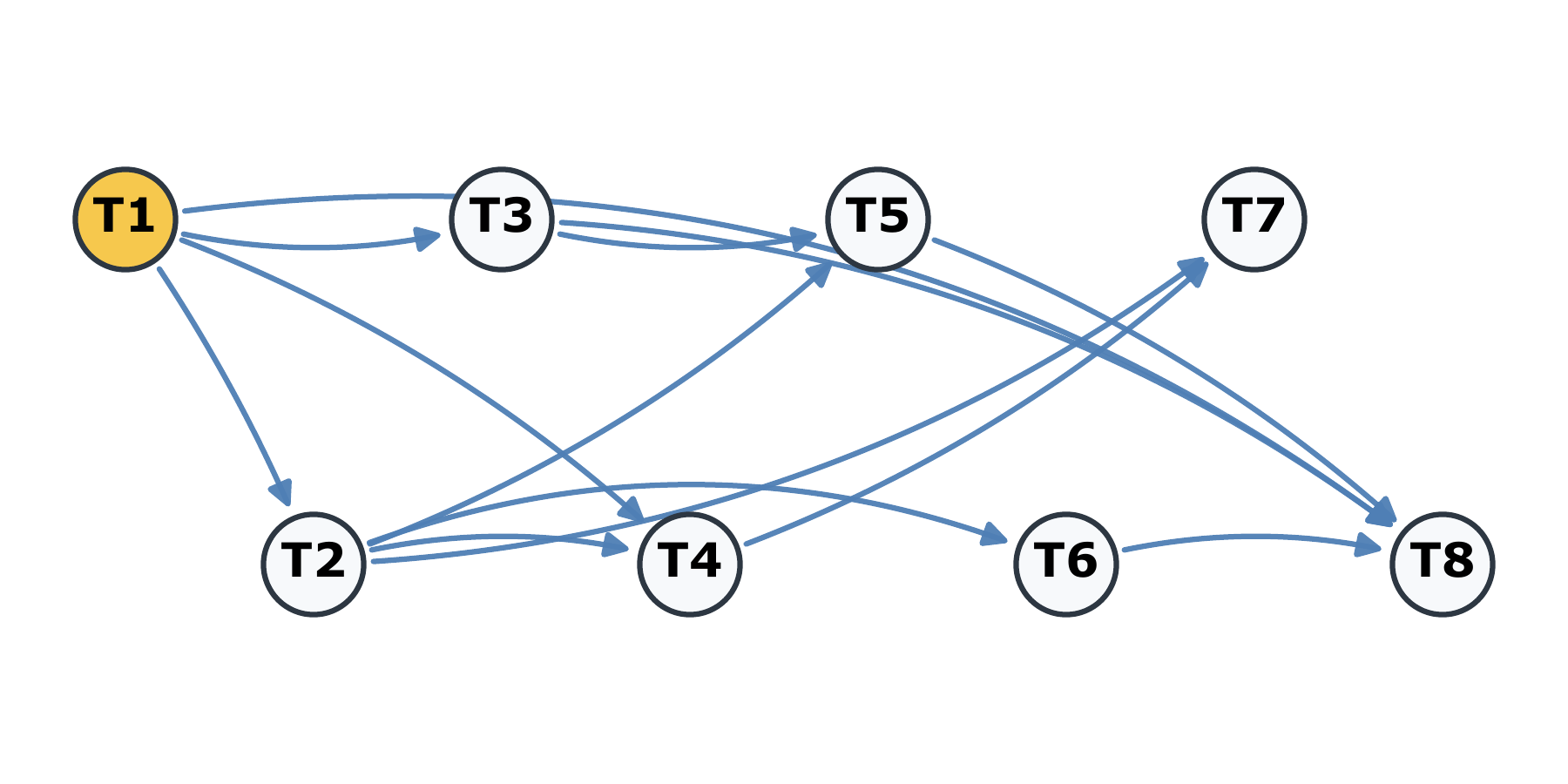}
\label{figTaskGraphC55}}
\subfloat[Complexity $0.65$]{\includegraphics[width=0.24\textwidth]{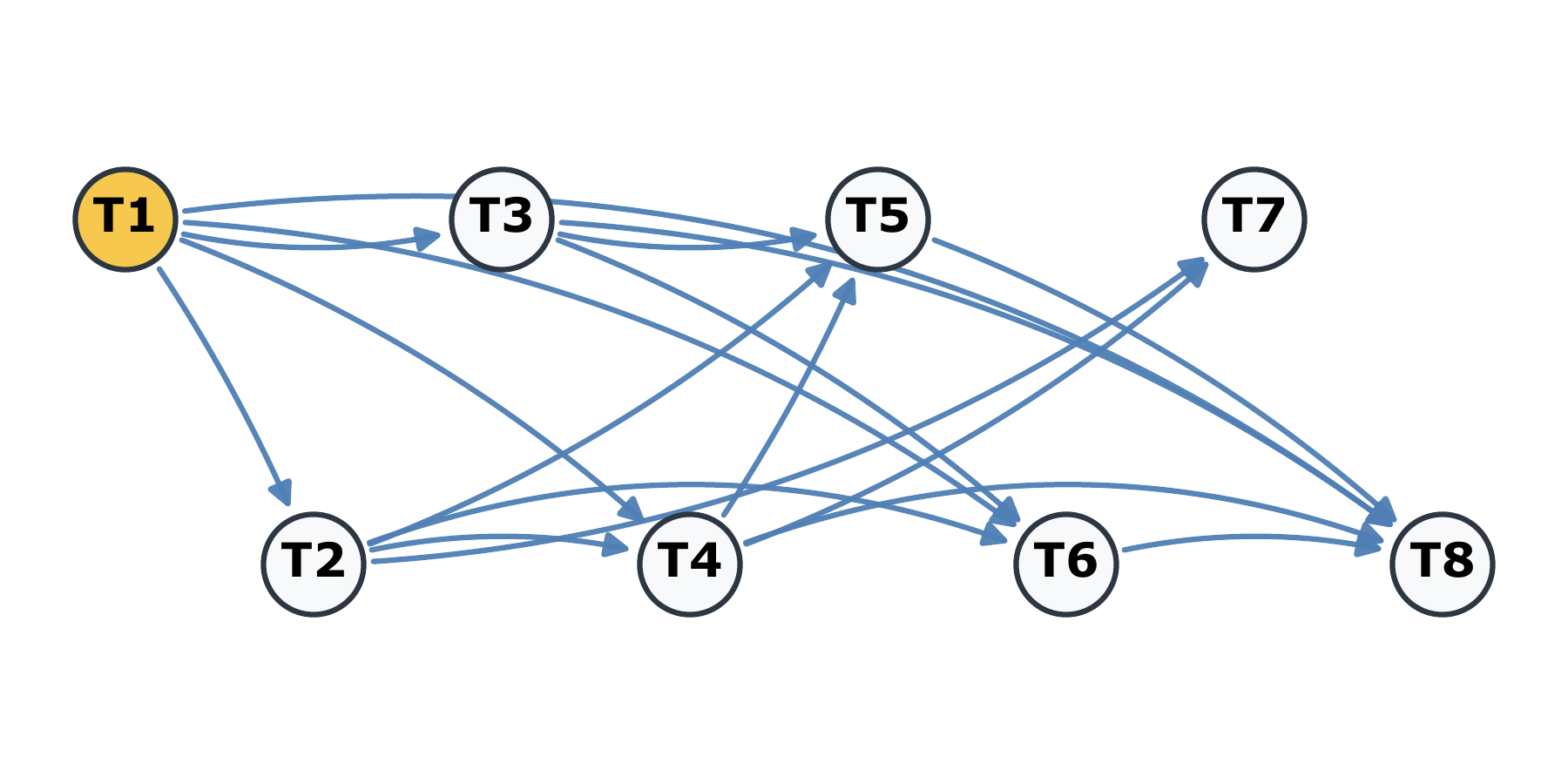}
\label{figTaskGraphC65}}
\subfloat[Complexity $0.75$]{\includegraphics[width=0.24\textwidth]{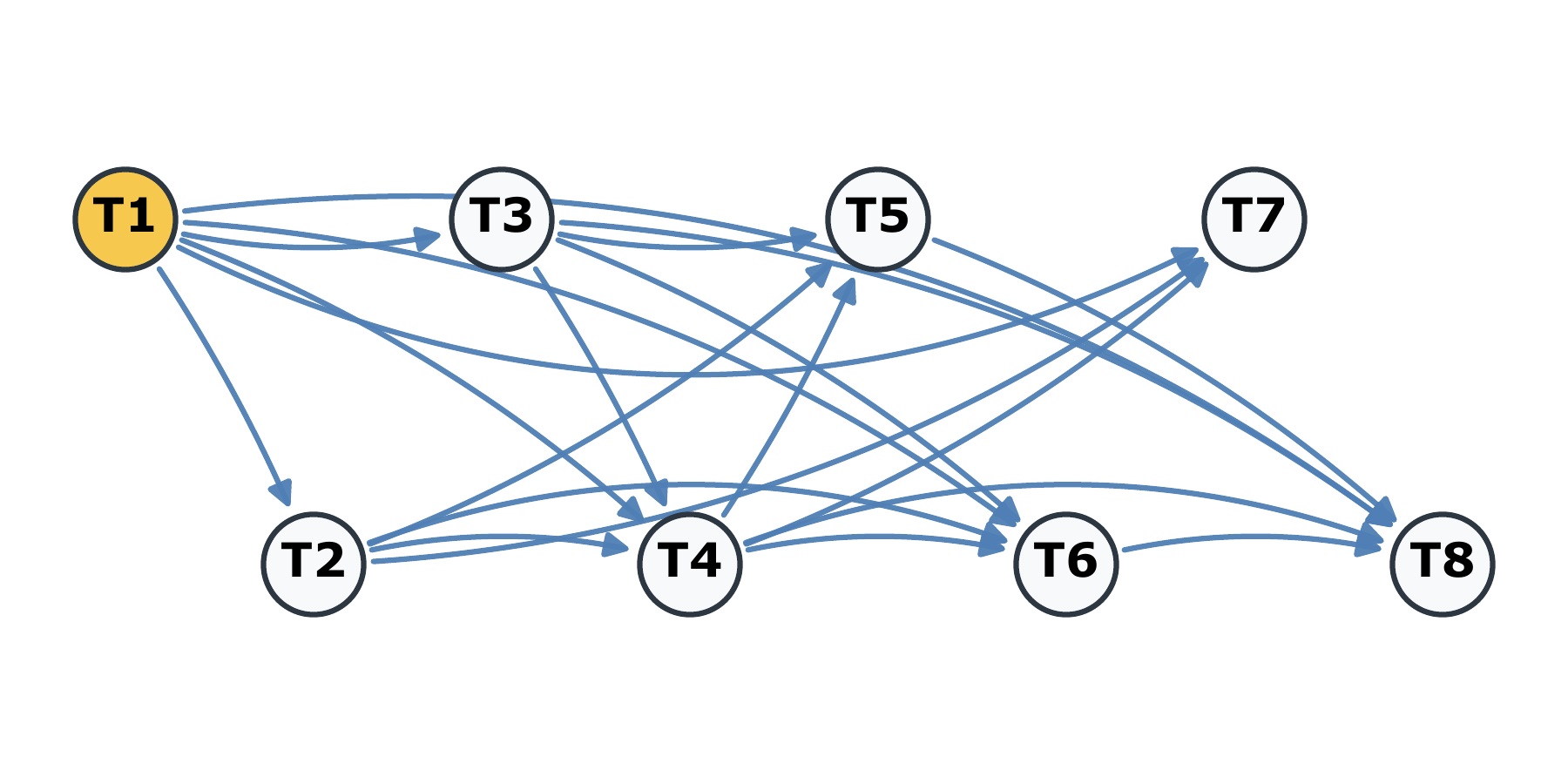}
\label{figTaskGraphC75}} \vspace{-6pt}
\caption{Task dependency graph examples with increasing graph complexity. Yellow nodes denote tasks without predecessor constraints and white nodes denote tasks constrained by predecessor relations.}
\label{figTaskGraphExamples}
\vspace{-12pt}
\end{figure*}

For task node $m\in\mathcal{M}$, its spatial attribute is denoted as $\mathbf{q}_m=[x_m,y_m,r_m]^{\top}$, where $\mathbf{q}_m^c=[x_m,y_m]^{\top}$ denotes the center of the task region and $r_m$ denotes the radius of the task region. The data size of task $m$ is denoted as $L_m$, the required central processing unit (CPU) cycles per bit is denoted as $c_m$, the task weight is denoted as $w_m$ and the deadline is denoted as $\tau_m^{\max}$. Therefore, the task node feature is defined as
\begin{equation}
\mathbf{h}^{\mathrm{Task}}_m=
\left[
\mathbf{q}_m^{\top},
L_m,
c_m,
w_m,
\tau_m^{\max},\delta_m^{\mathrm{act}},\delta_m^{\mathrm{cmp}}
\right]^{\top},
\label{eqTaskFeature}
\end{equation}
where $\delta_m^{\mathrm{act}}$ denotes whether task $m$ is active and $\delta_m^{\mathrm{cmp}}$ denotes whether task $m$ has been completed.
The logical dependency among tasks is represented by a task directed graph $\mathcal{G}_T=(\mathcal{M},\mathcal{E}_{\mathrm{TT}})$. If a directed edge $(p,m)\in\mathcal{E}_{\mathrm{TT}}$ exists, task $m$ can start only after task $p$ has been completed. The adjacency matrix of the task directed graph is denoted as $\mathbf{A}_{\mathrm{Task}}=[a^{\mathrm{Task}}_{pm}]\in\{0,1\}^{M\times M}$, where $a^{\mathrm{Task}}_{pm}=1$ indicates that task $p$ is a predecessor of task $m$ and $a^{\mathrm{Task}}_{pm}=0$ otherwise. The predecessor set of task $m$ is defined as
\begin{equation}
\mathcal{P}(m)=\{p\in\mathcal{M}\mid a^{\mathrm{Task}}_{pm}=1\}.
\label{eqPredecessorSet}
\end{equation}
To avoid cyclic execution dependencies, $\mathcal{G}_T$ is assumed to be a directed acyclic graph. Let $S_m$ denote the start time of task $m$ and $F_p$ denote the completion time of task $p$. The dependency constraint can be written as
\begin{equation}
S_m\geq F_p,\quad \forall p\in\mathcal{P}(m),\quad \forall m\in\mathcal{M}.
\label{eqDependencyConstraint}
\end{equation}
This constraint means that each task node becomes executable only after all its predecessor tasks have been completed, so the task graph can describe the execution order and logical relation in complex sensing tasks. The task complexity is defined as
\begin{equation}
\mathcal{C}_{T}
=
\frac{|\mathcal{E}_{\mathrm{TT}}|}{M(M-1)/2}
=
\frac{2}{M(M-1)}
\sum_{p\in\mathcal{M}}\sum_{m\in\mathcal{M}} a^{\mathrm{Task}}_{pm}.
\label{eqTaskComplexity}
\end{equation}
A larger $\mathcal{C}_{T}$ indicates a denser task dependency structure, implying stronger precedence constraints and lower potential parallelism among tasks. Task graphs of varying complexity are shown in Fig. \ref{figTaskGraphExamples}. 

For UAV node $i\in\mathcal{U}$, its two dimensional position is denoted as $\mathbf{p}_i=[x_i,y_i]^{\top}$ and its velocity is denoted as $\mathbf{v}_i$. The sensing rate of UAV $i$ is denoted as $\rho_i^{\mathrm{sen}}$, the maximum transmit power is denoted as $P_i^{\max}$, the maximum bandwidth is denoted as $B_i^{\max}$, the CPU frequency is denoted as $f_i$, the chip coefficient is denoted as $\kappa_i$, the maximum speed is denoted as $v_i^{\max}$ and the maximum acceleration is denoted as $a_i^{\max}$. Therefore, the UAV node feature is defined as
\begin{equation}
\mathbf{h}^{\mathrm{UAV}}_i=
\left[
\mathbf{p}_i^{\top},
\mathbf{v}_i^{\top},
\rho_i^{\mathrm{sen}},
P_i^{\max},
B_i^{\max},
f_i,
\kappa_i,
v_i^{\max},
a_i^{\max}
\right]^{\top}.
\label{eqUavFeature}
\end{equation}

The candidate communication relation among UAVs is represented by the adjacency matrix $\mathbf{A}=[a_{ij}]\in\{0,1\}^{N\times N}$, where $a_{ij}=1$ indicates that a communication link between UAV $i$ and UAV $j$ is retained and $a_{ij}=0$ otherwise. For UAV $i$, its neighbor set is defined as
\begin{equation}
\mathcal{N}_i=\{j\in\mathcal{U}\mid a_{ij}=1\}.
\label{eqNeighborSet}
\end{equation}
The edge set of the UAV communication graph is written as
\begin{equation}
\mathcal{E}_{\mathrm{UU}}=\{(i,j)\mid i,j\in\mathcal{U},a_{ij}=1\}.
\label{eqUavEdgeSet}
\end{equation}
The resulting UAV communication graph is denoted as $\mathcal{G}_U=(\mathcal{U},\mathcal{E}_{\mathrm{UU}})$. For the candidate communication edge $(i,j)\in\mathcal{E}_{\mathrm{UU}}$, the edge feature vector is defined as
\begin{equation}
\mathbf{e}^{\mathrm{UU}}_{ij}=
\left[
g_{ij},
\rho_{ij}^{\mathrm{LOS}}
\right]^{\top}.
\label{eqUavEdgeFeature}
\end{equation}
Here, $g_{ij}$ denotes the channel gain from UAV $i$ to UAV $j$ and $\rho_{ij}^{\mathrm{LOS}}$ denotes the line of sight probability of the corresponding link. To jointly consider channel strength and line of sight availability, the equivalent channel gain is defined as
\begin{equation}
\bar g_{ij}=g_{ij}\rho_{ij}^{\mathrm{LOS}}.
\label{eqEquivalentChannelGain}
\end{equation}
This edge feature reflects the link quality among UAVs and further affects communication topology selection, data offloading feasibility and cooperative computing efficiency. The task assignment matrix is denoted as $\mathbf{Y}=[y_{im}]\in\{0,1\}^{N\times M}$, where $y_{im}=1$ indicates that task $m$ is sensed by UAV $i$. 

\subsection{UAV Communication and Control Model}

For UAV $i\in\mathcal{U}$ and UAV $j\in\mathcal{U}$, the bandwidth allocation variable of communication link $(i,j)$ is denoted as $b_{ij}$ and the transmit power allocation variable is denoted as $p_{ij}$. The power allocation matrix is denoted as $\mathbf{P}=[p_{ij}]$ and the bandwidth allocation matrix is denoted as $\mathbf{B}=[b_{ij}]$. The noise power spectral density is denoted as $N_0$. Based on the equivalent channel gain $\bar g_{ij}$ defined above, the achievable communication rate of link $(i,j)$ is defined as
\begin{equation}
R_{ij}
=
b_{ij}\log_2
\left(
1+\frac{p_{ij}\bar g_{ij}}{N_0b_{ij}}
\right).
\label{eqCommRate}
\end{equation}
Here, $R_{ij}$ denotes the link rate when UAV $i$ transmits data to UAV $j$. The communication resource of UAV $i$ is constrained by its own resource budget. The power and bandwidth occupied by all outgoing communication links of UAV $i$ satisfy
\begin{equation}
\begin{aligned}
\sum_{j\in\mathcal{U}}p_{ij}
&\leq P_i^{\max},\\
\sum_{j\in\mathcal{U}}b_{ij}
&\leq B_i^{\max},
\quad \forall i\in\mathcal{U}.
\end{aligned}
\label{eqCommBudget}
\end{equation}

The link resource variables also need to be consistent with the communication adjacency matrix $\mathbf{A}$. 
If a communication link is retained, it needs to satisfy the minimum rate constraint. The minimum rate requirement is denoted as $R^{\min}$ and we have
\begin{equation}
R_{ij}
\geq
a_{ij}R^{\min},
\quad
\forall i,j\in\mathcal{U}.
\label{eqMinRate}
\end{equation}
The motion state of each UAV is described by a double integrator model. The acceleration control input is denoted as $\mathbf{u}_i$. The kinematic equations are defined as $\dot{\mathbf{p}}_i=\mathbf{v}_i, \dot{\mathbf{v}}_i=\mathbf{u}_i,
\forall i\in\mathcal{U}$.
The reference position of UAV $i$ is denoted as $\mathbf{p}_i^{\mathrm{ref}}$ and the reference velocity is denoted as $\mathbf{v}_i^{\mathrm{ref}}$. The proportional control gain is denoted as $k_p$ and the derivative control gain is denoted as $k_v$. The nominal control input obtained by the proportional derivative controller is defined as
\begin{equation}
\mathbf{u}_i^0
=
k_p\left(\mathbf{p}_i^{\mathrm{ref}}-\mathbf{p}_i\right)
+
k_v\left(\mathbf{v}_i^{\mathrm{ref}}-\mathbf{v}_i\right).
\label{eqNominalControl}
\end{equation}
Here, $\mathbf{u}_i^0$ denotes the control input before considering safety constraints. The motion constraints are given by $
\left\|\mathbf{v}_i\right\|_2\leq v_i^{\max},\left\|\mathbf{u}_i\right\|_2 \leq a_i^{\max},\forall i\in\mathcal{U}$.

To maintain the safety distance among UAVs, the minimum safety distance is denoted as $d_{\min}$. For each neighboring UAV $j\in\mathcal{N}_i$, the safety function is defined as
\begin{equation}
h_{ij}=\left\|\mathbf{p}_i-\mathbf{p}_j\right\|_2^2-d_{\min}^2.
\label{eqSafetyFunction}
\end{equation}
Here, $h_{ij}\geq 0$ means that the distance between UAV $i$ and UAV $j$ is not smaller than $d_{\min}$. The control barrier function constraint is defined as
\begin{equation}
\ddot h_{ij}
+
k_{h,1}\dot h_{ij}
+
k_{h,0}h_{ij}
\geq
0,
\quad
\forall j\in\mathcal{N}_i.
\label{eqCbfConstraint}
\end{equation}
Here, $k_{h,0}>0$ and $k_{h,1}>0$ denote positive control parameters in the safety constraint. Given the nominal control input $\mathbf{u}_i^0$, the applied control input $\mathbf{u}_i^\star$ is obtained through constrained projection. This projection problem is defined as
\begin{equation}
\begin{aligned}
\mathbf{u}_i^\star
=&
\arg\min_{\mathbf{u}_i}
\left\|\mathbf{u}_i-\mathbf{u}_i^0\right\|_2^2\\
\mathrm{s.t.}\quad
&
\left\|\mathbf{u}_i\right\|_2\leq a_i^{\max},\\
&
\left\|\mathbf{v}_i\right\|_2\leq v_i^{\max},\\
&
\ddot h_{ij}
+
k_{h,1}\dot h_{ij}
+
k_{h,0}h_{ij}
\geq
0,
\quad \forall j\in\mathcal{N}_i,
\end{aligned}
\label{eqControlProjection}
\end{equation}
where $\mathbf{u}_i^\star$ denotes the control input that satisfies the motion constraints and the safety distance constraints.

\subsection{Task Completion Time}

Let $\mathcal{M}_{\mathrm{act}}$ denote the set of unfinished tasks and $\chi_m$ denote the completion state of task $m$. If $\chi_m=1$, task $m$ has been completed. Otherwise, it remains unfinished. At each execution time, only tasks whose predecessors have all been completed can be scheduled, and the ready task set is given by
\begin{equation}
\mathcal{M}_{\mathrm{ready}}
=
\{m\in\mathcal{M}_{\mathrm{act}}\mid \chi_p=1,\forall p\in\mathcal{P}(m)\}.
\label{eqReadyTaskSetEn}
\end{equation}
For each task $m$, the assignment variable $y_{im}\in\{0,1\}$ indicates whether task $m$ is sensed by UAV $i$, with $\sum_{i\in\mathcal{U}}y_{im}=1$. Let $\tilde{\mathbf{p}}_i$ be the current position of UAV $i$, $\mathbf{q}_m^c$ be the center of task region $m$, and $\bar v_i$ be the average flight speed. The target waypoint $\mathbf{q}_{i\to m}$ is the intersection point between the circular task boundary and the line segment from $\mathbf{p}_i(t)$ to $\mathbf{q}_m^c$.
The flight time from UAV $i$ to task $m$ is $T_{i,m}^{\mathrm{fly}}=\|\tilde{\mathbf{p}}_i-\mathbf{q}_{i\to m}\|_2/\bar v_i$, and the task flight time is $T_m^{\mathrm{fly}}=\sum_{i\in\mathcal{U}}y_{im}T_{i,m}^{\mathrm{fly}}$.
Given the data size $L_m$ and the sensing rate $\rho_i^{\mathrm{sen}}$, the sensing collection time is $T_m^{\mathrm{sen}}=\sum_{i\in\mathcal{U}}y_{im}L_m/\rho_i^{\mathrm{sen}}$.

Let $\Omega_i^{\mathrm{sen}}$ denote the available time of sensing UAV $i$, and let $F_p$ denote the completion time of predecessor task $p$. The start time of task $m$ is determined by both UAV availability and task precedence, which is written as
\begin{equation}
S_m
=
\sum_{i\in\mathcal{U}}y_{im}
\max\left\{
\Omega_i^{\mathrm{sen}},
\max_{p\in\mathcal{P}(m)}F_p
\right\}.
\label{eqTaskStartTimeEn}
\end{equation}
If task $m$ has no predecessor, $\max_{p\in\mathcal{P}(m)}F_p$ is set to $0$. The sensing finish time is then $C_m^{\mathrm{sen}}=S_m+T_m^{\mathrm{fly}}+T_m^{\mathrm{sen}}$. After sensing is completed, the position of the assigned UAV is updated to the target position $\mathbf{q}_{i\to m}$.

After the sensing stage, task data are split among feasible computing UAVs. For task $m$ assigned to sensing UAV $i$, let $\mathcal{C}_{i,m}$ denote the feasible computing set, which includes the local UAV and available computing neighbors. The computation splitting tensor is denoted as $\mathbf{X}=[x_{ji,m}]$, where $x_{ji,m}$ denotes the fraction of task $m$ data collected by UAV $i$ and computed by UAV $j$. Since all sensing data should be processed by feasible computing UAVs, the splitting ratios satisfy
\begin{equation}
\begin{aligned}
\sum_{j\in\mathcal{C}_{i,m}}x_{ji,m}
&=y_{im},
\quad
\forall i\in\mathcal{U},\ \forall m\in\mathcal{M},
\\
x_{ji,m}
&\geq0,
\quad
\forall j\in\mathcal{C}_{i,m}.
\end{aligned}
\label{eqSplitRatioConstraint}
\end{equation}

The amount of data transmitted from UAV $i$ to UAV $j$ is then $D_{ji,m}=x_{ji,m}L_m$.

For a remote computing block, the transmission time is $T_{ji,m}^{\mathrm{tx}}=D_{ji,m}/R_{ij}$. For a local computing block, $T_{ii,m}^{\mathrm{tx}}=0$. Therefore, the data arrival time at computing UAV $j$ is $T_{ji,m}^{\mathrm{data}}=C_m^{\mathrm{sen}}+T_{ji,m}^{\mathrm{tx}}$.

Each computing UAV maintains an independent first come first served computation queue. Let $\Omega_j^{\mathrm{cmp}}$ denote the current available time of computing UAV $j$. The computation start time of data block $(j,i,m)$ is $B_{ji,m}^{\mathrm{cmp}}=\max\{T_{ji,m}^{\mathrm{data}},\Omega_j^{\mathrm{cmp}}\}$.
Given the CPU frequency $f_j$ and the required CPU cycles per bit $c_m$, the computation time is $T_{ji,m}^{\mathrm{cmp}}=\kappa_{j} D_{ji,m}c_m/f_j$, and the block finish time is $F_{ji,m}^{\mathrm{blk}}=B_{ji,m}^{\mathrm{cmp}}+T_{ji,m}^{\mathrm{cmp}}$. After the block is computed, the available time of computing UAV $j$ is updated as $\Omega_j^{\mathrm{cmp}}\leftarrow F_{ji,m}^{\mathrm{blk}}$.

Task $m$ is regarded as completed only after all its data blocks have been computed. Its completion time is
\begin{equation}
F_m
=
\sum_{i\in\mathcal{U}}y_{im}
\max_{\substack{j\in\mathcal{C}_{i,m}\\ x_{ji,m}>0}}
F_{ji,m}^{\mathrm{blk}}.
\label{eqTaskFinishTimeSystemEn}
\end{equation}
Given the deadline $\tau_m^{\max}$, the completion indicator is defined as
\begin{equation}
\chi_m
=
\begin{cases}
1,\quad F_m\leq \tau_m^{\max},\\
0,\quad F_m>\tau_m^{\max}.
\end{cases}
\label{eqCompletionIndicatorSystemEn}
\end{equation}
The total completion time of the system is determined by the last completed task, namely $T^{\mathrm{total}}=\max_{m\in\mathcal{M}}F_m$.
\section{Proposed Solution Framework}

\subsection{Problem Formulation}
Based on the above system model, UAV sensing assignment, communication topology generation, computation splitting and communication resource allocation are jointly formulated as a mixed integer nonlinear optimization problem.
The optimization objective is to maximize the number of successfully completed tasks and minimize the total task completion time. The joint optimization problem can be written as

\begin{subequations}
\label{eqPzeroEnglish}
\begin{align}
\mathbf{P0}\quad
&\max_{\mathbf{Y},\mathbf{A},\mathbf{X},\mathbf{P},\mathbf{B}}\quad
 \eta_1\sum_{m\in\mathcal{M}}w_m\chi_m-\eta_2T^{\mathrm{total}},
\label{eqPzeroObjEnglish}
\\
\mathrm{s.t.}\quad
& \sum_{i\in\mathcal{U}}y_{im}=1,
\quad \forall m\in\mathcal{M}_{\mathrm{act}},
\label{eqPzeroAssignEnglish}
\\
& \lambda_2(\mathbf{L})>0,
\label{eqPzeroTopoEnglish}
\\
& 0\leq p_{ij}\leq P_i^{\max},
\quad 0\leq b_{ij}\leq B_i^{\max},
\quad \forall i,j\in\mathcal{U},
\label{eqPzeroBoundEnglish}
\\
& R_{ij}\geq R^{\min},
\quad \forall i,j\in\mathcal{U},
\label{eqPzeroRateEnglish}
\\
& \sum_{j\in\mathcal{U}}p_{ij}\leq P_i^{\max},
\quad \sum_{j\in\mathcal{U}}b_{ij}\leq B_i^{\max},
\quad \forall i\in\mathcal{U}.
\label{eqPzeroResourceEnglish}
\end{align}
\end{subequations}
where $\eta_1$ and $\eta_2$ are nonnegative weights. \eqref{eqPzeroAssignEnglish} indicates that each active task is assigned to exactly one sensing UAV. \eqref{eqPzeroTopoEnglish} ensures the connectivity of the retained communication topology. \eqref{eqPzeroBoundEnglish} gives the link level power and bandwidth bounds. \eqref{eqPzeroRateEnglish} imposes the minimum transmission rate requirement for each communication link. \eqref{eqPzeroResourceEnglish} constrains the total transmit power and bandwidth consumed by each UAV within its available resource budgets.

\subsection{GAT Based Feature Extraction}

The heterogeneous graph encoder maps UAV attributes, task attributes, edge attributes and execution states into task oriented embeddings using GAT. For UAV $i\in\mathcal{U}$, the original UAV feature is denoted as $\mathbf{h}^{\mathrm{UAV}}_i$.
For task $m\in\mathcal{M}$, the initial task embedding is denoted as $\mathbf{h}^{(0)\mathrm{Task}}_m=\phi_T^{0}(\mathbf{h}^{\mathrm{Task}}_m)$. The initial UAV embedding is denoted as $\mathbf{h}^{(0)\mathrm{UAV}}_i=\phi_U^{0}(\mathbf{h}^{\mathrm{UAV}}_i)$.
Here, $\phi_U^{0}(\cdot)$ and $\phi_T^{0}(\cdot)$ are trainable node encoders.

For the UAV communication graph, the neighbor set of UAV $i$ is denoted as $\mathcal{N}_i$, and the edge feature between UAV $i$ and UAV $j$ is denoted as $\mathbf{e}^{\mathrm{UU}}_{ij}$. At message passing layer $r$, the edge aware attention score from UAV $j$ to UAV $i$ is computed by
\begin{equation}
s_{ij}^{(r)}
=
\varphi_U^{(r)}
\left(
\mathbf{h}^{(r)\mathrm{UAV}}_i,
\mathbf{h}^{(r)\mathrm{UAV}}_j,
\mathbf{e}^{\mathrm{UU}}_{ij}
\right),
\quad
j\in\mathcal{N}_i .
\label{eqUavAttentionScore}
\end{equation}
The corresponding attention coefficient is defined as
\begin{equation}
\alpha_{ij}^{(r)}
=
\frac{\exp\left(s_{ij}^{(r)}\right)}
{\sum_{k\in\mathcal{N}_i}\exp\left(s_{ik}^{(r)}\right)},
\quad
j\in\mathcal{N}_i ,
\label{eqUavAttentionCoefficient}
\end{equation}
where $\varphi_U^{(r)}(\cdot)$ is a trainable attention scoring function. Based on the attention coefficients, the UAV embedding is updated by aggregating the transformed neighbor features as
\begin{equation}
\begin{aligned}
\boldsymbol{\mu}_i^{(r)}
&=
\sum_{j\in\mathcal{N}_i}
\alpha_{ij}^{(r)}
\mathbf{W}_{U}^{(r)}
\mathbf{h}^{(r)\mathrm{UAV}}_j,
\\
\mathbf{h}^{(r+1)\mathrm{UAV}}_i
&=
\sigma_U^{(r)}
\left(
\mathbf{W}_{0}^{(r)}
\mathbf{h}^{(r)\mathrm{UAV}}_i
+
\boldsymbol{\mu}_i^{(r)}
\right),
\\
&\quad
\forall i\in\mathcal{U},
\end{aligned}
\label{eqUavEmbeddingUpdate}
\end{equation}
where $\mathbf{W}_{0}^{(r)}$ and $\mathbf{W}_{U}^{(r)}$ are trainable matrices, and $\sigma_U^{(r)}(\cdot)$ denotes a nonlinear activation function. Through this edge aware attention aggregation, each UAV embedding is updated according to both the states of its neighboring UAVs and the quality of the corresponding communication links.

For the directed task graph, the predecessor set and successor set of task $m$ are denoted as $\mathcal{P}(m)$ and $\mathcal{S}(m)$, respectively. At layer $r$, the predecessor message and successor message of task $m$ are defined as
\begin{equation}
\begin{aligned}
\boldsymbol{\nu}_{m,\mathrm{pred}}^{(r)}
&=
\frac{1}{|\mathcal{P}(m)|+\epsilon}
\sum_{p\in\mathcal{P}(m)}
\mathbf{W}_{p}^{(r)}
\mathbf{h}^{(r)\mathrm{Task}}_p,
\\
\boldsymbol{\nu}_{m,\mathrm{succ}}^{(r)}
&=
\frac{1}{|\mathcal{S}(m)|+\epsilon}
\sum_{s\in\mathcal{S}(m)}
\mathbf{W}_{s}^{(r)}
\mathbf{h}^{(r)\mathrm{Task}}_s ,
\end{aligned}
\label{eqTaskDirectedMessage}
\end{equation}
where $\epsilon$ is a small positive constant, and $\mathbf{W}_{p}^{(r)}$ and $\mathbf{W}_{s}^{(r)}$ are trainable matrices. The task embedding is updated as
\begin{equation}
\begin{aligned}
\mathbf{h}^{(r+1)\mathrm{Task}}_m
&=
\mathbf{h}^{(r)\mathrm{Task}}_m
+
\phi_T^{(r)}
\left(
\left[
\begin{array}{c}
\mathbf{h}^{(r)\mathrm{Task}}_m\\
\boldsymbol{\nu}_{m,\mathrm{pred}}^{(r)}\\
\boldsymbol{\nu}_{m,\mathrm{succ}}^{(r)}
\end{array}
\right]
\right),
\\
&\quad
\forall m\in\mathcal{M},
\end{aligned}
\label{eqTaskEmbeddingUpdate}
\end{equation}
where $\phi_T^{(r)}(\cdot)$ is a trainable task update function.

After the intra graph message passing, all UAV nodes and task nodes are placed into a unified node set. The type embeddings of UAV nodes and task nodes are denoted as $\mathbf{t}_U$ and $\mathbf{t}_T$, respectively. The unified node representation is initialized as
\begin{equation}
\begin{aligned}
\mathbf{Z}^{(0)}
&=
\left[
\begin{array}{c}
\mathbf{h}^{(R)\mathrm{UAV}}_1+\mathbf{t}_U,
\ldots,
\mathbf{h}^{(R)\mathrm{UAV}}_N+\mathbf{t}_U\\
\mathbf{h}^{(R)\mathrm{Task}}_1+\mathbf{t}_T,
\ldots,
\mathbf{h}^{(R)\mathrm{Task}}_M+\mathbf{t}_T
\end{array}
\right]^{\top},
\end{aligned}
\label{eqAllNodeInput}
\end{equation}
where $R$ is the number of intra graph message passing layers. For the all node interaction layer $l$, the unified node representation is updated as
\begin{equation}
\widetilde{\mathbf{Z}}^{(l+1)}
=
\mathbf{Z}^{(l)}
+
\mathcal{A}^{(l)}
\left(
\mathbf{Z}^{(l)}
\right),
\label{eqAllNodeAttention}
\end{equation}
\begin{equation}
\mathbf{Z}^{(l+1)}
=
\widetilde{\mathbf{Z}}^{(l+1)}
+
\mathcal{F}^{(l)}
\left(
\widetilde{\mathbf{Z}}^{(l+1)}
\right),
\label{eqAllNodeFfn}
\end{equation}
where $\mathcal{A}^{(l)}(\cdot)$ denotes a trainable multi head attention mapping, and $\mathcal{F}^{(l)}(\cdot)$ denotes a trainable feed forward transformation. The refined UAV embeddings and task embeddings extracted from $\mathbf{Z}^{(L)}$ are denoted as $\bar{\mathbf{h}}^{\mathrm{UAV}}_i$ and $\bar{\mathbf{h}}^{\mathrm{Task}}_m$, respectively.

\subsection{Task Allocation and Topology Decision}

Based on the heterogeneous graph embeddings, the policy network jointly generates the task assignment matrix $\mathbf{Y}$, the communication topology matrix $\mathbf{A}$ and the computation splitting tensor $\mathbf{X}$. Let $\mathbf{z}_i^{U}$, $\mathbf{z}_m^{T}$ and $\mathbf{z}_{G}$ denote the final UAV embedding, task embedding and graph level embedding, respectively. The task assignment and topology decisions are produced from the same graph state, while the computation splitting decision is conditioned on the generated assignment and topology.

For each active task $m\in\mathcal{M}_{\mathrm{act}}$, the assignment probability is defined as
\begin{equation}
\pi_{\theta}^{Y}
\left(
y_{im}=1\mid\mathcal{G}
\right)
=
\frac{
\exp
\left(
\phi_Y
\left(
\mathbf{z}_i^{U},
\mathbf{z}_m^{T},
\mathbf{z}_{G}
\right)
\right)
}
{
\sum_{k\in\mathcal{U}}
\exp
\left(
\phi_Y
\left(
\mathbf{z}_k^{U},
\mathbf{z}_m^{T},
\mathbf{z}_{G}
\right)
\right)
},
\quad
i\in\mathcal{U},
\label{eqAssignmentProb}
\end{equation}
where $\phi_Y(\cdot)$ is a trainable assignment scoring function. The sampled assignment satisfies $y_{im}\in\{0,1\}$ and $\sum_{i\in\mathcal{U}}y_{im}=1$ for each $m\in\mathcal{M}_{\mathrm{act}}$.

For each UAV pair $(i,j)$, the topology probability is given by
\begin{equation}
\begin{aligned}
\pi_{\theta}^{A}
\left(
a_{ij}=1\mid\mathcal{G}
\right)
&=
a_{ij}^{\mathrm{cand}}
\sigma
\left(
\phi_A
\left(
\mathbf{z}_i^{U},
\mathbf{z}_j^{U},
\mathbf{e}^{\mathrm{UU}}_{ij},
\mathbf{z}_{G}
\right)
\right),
\\
&\quad
i,j\in\mathcal{U},
\end{aligned}
\label{eqTopologyProb}
\end{equation}
where $\phi_A(\cdot)$ is a trainable topology scoring function and $a_{ij}^{\mathrm{cand}}$ indicates whether link $(i,j)$ belongs to the candidate communication graph. The generated topology satisfies $a_{ij}\in\{0,1\}$, $a_{ij}\leq a_{ij}^{\mathrm{cand}}$, $a_{ij}=a_{ji}$, $a_{ii}=0$ and $\lambda_2(\mathbf{L})>0$.

Conditioned on $\mathbf{Y}$ and $\mathbf{A}$, the computation splitting head produces the data splitting ratio. For task $m$ assigned to sensing UAV $i$, let $\mathcal{C}_{i,m}$ denote the feasible computing set, which includes UAV $i$ itself and its available computing neighbors. For $j\in\mathcal{C}_{i,m}$, the splitting ratio is obtained by a masked softmax as
\begin{equation}
\begin{aligned}
x_{ji,m}
&=
\frac{
\exp
\left(
\phi_X
\left(
\mathbf{z}_i^{U},
\mathbf{z}_j^{U},
\mathbf{z}_m^{T},
\mathbf{z}_{G},
\mathbf{e}^{\mathrm{UU}}_{ij}
\right)
\right)
}
{
\sum_{r\in\mathcal{C}_{i,m}}
\exp
\left(
\phi_X
\left(
\mathbf{z}_i^{U},
\mathbf{z}_r^{U},
\mathbf{z}_m^{T},
\mathbf{z}_{G},
\mathbf{e}^{\mathrm{UU}}_{ir}
\right)
\right)
},
\\
&\quad
j\in\mathcal{C}_{i,m},
\end{aligned}
\label{eqSplitProb}
\end{equation}
where $\phi_X(\cdot)$ is a trainable splitting scoring function. For local computing, $\mathbf{e}^{\mathrm{UU}}_{ii}$ is set as a zero vector. For illegal computing nodes, $x_{ji,m}=0$. The splitting decision satisfies $x_{ji,m}\geq0$ and $\sum_{j\in\mathcal{C}_{i,m}}x_{ji,m}=y_{im}$ for all $i\in\mathcal{U}$ and $m\in\mathcal{M}_{\mathrm{act}}$.

The complete policy is factorized as
\begin{equation}
\begin{aligned}
\pi_{\theta}
\left(
\mathbf{Y},\mathbf{A},\mathbf{X}
\mid
\mathcal{G}
\right)
&=
\pi_{\theta}^{Y}
\left(
\mathbf{Y}\mid\mathcal{G}
\right)
\pi_{\theta}^{A}
\left(
\mathbf{A}\mid\mathcal{G}
\right)
\\
&\quad\times
\pi_{\theta}^{X}
\left(
\mathbf{X}\mid\mathcal{G},\mathbf{Y},\mathbf{A}
\right),
\end{aligned}
\label{eqJointPolicyFactor}
\end{equation}
where $\theta$ denotes all trainable policy parameters. In this way, $\mathbf{Y}$ determines the sensing UAV for each active task, $\mathbf{A}$ determines the retained communication topology and $\mathbf{X}$ determines the computation splitting ratio under the selected assignment and topology.

\subsection{Communication Resource Allocation}
After the task assignment matrix $\mathbf{Y}$, the communication topology matrix $\mathbf{A}$ and the computation splitting tensor $\mathbf{X}$ are generated, the communication resource allocation module determines the bandwidth matrix $\mathbf{B}=[b_{ij}]$ and the power matrix $\mathbf{P}=[p_{ij}]$ without changing $\mathbf{Y}$, $\mathbf{A}$ or $\mathbf{X}$. The amount of task data carried by link $(i,j)$ is given by
\begin{equation}
D_{ij}
=
\sum_{m\in\mathcal{M}}D_{ji,m}.
\label{eqLinkDataAmount}
\end{equation}
Only retained directed links are considered for resource allocation, namely
\begin{equation}
\mathcal{E}^{\mathrm{tx}}
=
\{(i,j)\mid a_{ij}=1,\ i\neq j\}.
\label{eqActiveTxLinkSet}
\end{equation}

For link $(i,j)\in\mathcal{E}^{\mathrm{tx}}$, the achievable transmission rate is
\begin{equation}
R_{ij}
=
b_{ij}\log_2
\left(
1+
\frac{p_{ij}\bar g_{ij}}{N_0b_{ij}}
\right),
\label{eqScaCommRate}
\end{equation}
where $\bar g_{ij}$ is the equivalent channel gain and $N_0$ is the noise power spectral density. To satisfy the minimum rate requirement $R_{ij}\geq R^{\min}$, the constraint can be equivalently written as a lower bound on the transmit power
\begin{equation}
p_{ij}
\geq
p_{ij}^{\min}(b_{ij})
=
\left(
2^{\frac{R^{\min}}{b_{ij}}}
-
1
\right)
\frac{N_0b_{ij}}{\bar g_{ij}}.
\label{eqPowerLowerBound}
\end{equation}
Since $p_{ij}^{\min}(b_{ij})$ is nonlinear with respect to $b_{ij}$, successive convex approximation (SCA) is used to obtain a tractable approximation. At iteration $q$, $p_{ij}^{\min}(b_{ij})$ is linearized around the current point $b_{ij}^{(q)}$ as
\begin{equation}
\widehat p_{ij}^{\min,(q)}(b_{ij})
=
p_{ij}^{\min}(b_{ij}^{(q)})
+
d_{ij}^{(q)}
\left(
b_{ij}-b_{ij}^{(q)}
\right),
\label{eqLinearizedPowerBound}
\end{equation}
where $d_{ij}^{(q)}=\left.\frac{\partial p_{ij}^{\min}(b)}{\partial b}\right|_{b=b_{ij}^{(q)}}$.

Under this approximation, the SCA subproblem at iteration $q$ is formulated as
\begin{equation}
\begin{aligned}
\min_{\mathbf{P},\mathbf{B}}\quad
& \sum_{(i,j)\in\mathcal{E}^{\mathrm{tx}}} p_{ij}
\\
\mathrm{s.t.}\quad
& b_{ij}\geq0,\quad p_{ij}\geq0,\quad \forall (i,j)\in\mathcal{E}^{\mathrm{tx}},
\\
& p_{ij}\geq \widehat p_{ij}^{\min,(q)}(b_{ij}),
\quad \forall (i,j)\in\mathcal{E}^{\mathrm{tx}},
\\
& \sum_{j:(i,j)\in\mathcal{E}^{\mathrm{tx}}}b_{ij}\leq B_i^{\max},
\\
&
\sum_{j:(i,j)\in\mathcal{E}^{\mathrm{tx}}}p_{ij}\leq P_i^{\max},
\quad \forall i\in\mathcal{U}.
\end{aligned}
\label{eqScaSubproblem}
\end{equation}
This subproblem has a linear objective and affine constraints, and can be solved efficiently. To improve numerical stability, the bandwidth variable is updated by
\begin{equation}
b_{ij}^{(q+1)}
=
\rho b_{ij}^{\star}
+
(1-\rho)b_{ij}^{(q)},
\quad
0<\rho\leq1,
\label{eqDampingUpdate}
\end{equation}
where $b_{ij}^{\star}$ is the solution of \eqref{eqScaSubproblem}. The SCA procedure is repeated until convergence or the maximum number of iterations is reached.

After the SCA based baseline allocation, the residual resources of UAV $i$ are
\begin{equation}
B_i^{\mathrm{res}}
=
B_i^{\max}
-
\sum_{j:(i,j)\in\mathcal{E}^{\mathrm{tx}}}b_{ij},
\quad
P_i^{\mathrm{res}}
=
P_i^{\max}
-
\sum_{j:(i,j)\in\mathcal{E}^{\mathrm{tx}}}p_{ij}.
\label{eqResidualResourceSca}
\end{equation}
The residual resources are further assigned to data carrying links
$\mathcal{E}^{\mathrm{data}}=\{(i,j)\mid (i,j)\in\mathcal{E}^{\mathrm{tx}},D_{ij}>0\}$.
For $(i,j)\in\mathcal{E}^{\mathrm{data}}$, the enhancement weight is defined as
\begin{equation}
w_{ij}=D_{ij}\bar g_{ij}.
\label{eqEnhancementWeight}
\end{equation}
The additional bandwidth and power are allocated by
\begin{equation}
\begin{aligned}
\Delta b_{ij}
&=
B_i^{\mathrm{res}}
\frac{w_{ij}}{\sum_{k:(i,k)\in\mathcal{E}^{\mathrm{data}}}w_{ik}+\epsilon},
\\
\Delta p_{ij}
&=
P_i^{\mathrm{res}}
\frac{w_{ij}}{\sum_{k:(i,k)\in\mathcal{E}^{\mathrm{data}}}w_{ik}+\epsilon}.
\end{aligned}
\label{eqResidualDistribution}
\end{equation}
The final resource allocation is updated as
\begin{equation}
b_{ij}\leftarrow b_{ij}+\Delta b_{ij},
\quad
p_{ij}\leftarrow p_{ij}+\Delta p_{ij},
\quad
(i,j)\in\mathcal{E}^{\mathrm{data}}.
\label{eqFinalResourceSca}
\end{equation}
Given the final $b_{ij}$ and $p_{ij}$, the transmission time of data block $(j,i,m)$ is computed by $T_{ji,m}^{\mathrm{tx}}=D_{ji,m}/R_{ij}$. For local computing, $T_{ii,m}^{\mathrm{tx}}=0$.

\subsection{Prediction and Decision Reconfiguration}
The completion time prediction module is designed to evaluate candidate decisions under the current system state. At decision step $k$, the current execution state is denoted as $\mathcal{S}_k$, and the remaining active task set is denoted as $\mathcal{M}_k^{\mathrm{rem}}$. A candidate decision is represented by
\begin{equation}
\mathcal{D}_k
=
\left\{
\mathbf{Y}_k,
\mathbf{A}_k,
\mathbf{X}_k
\right\},
\label{eqCandidateDecision}
\end{equation}
where $\mathbf{Y}_k$, $\mathbf{A}_k$ and $\mathbf{X}_k$ denote the task assignment, communication topology and computation splitting decisions for the remaining tasks, respectively.

The prediction head takes the graph embeddings, candidate decision and queue states as inputs. Let $\boldsymbol{\xi}_k$ denote the decision dependent structural feature extracted from the current state, which is written as
\begin{equation}
\boldsymbol{\xi}_k
=
f_{\xi}
\left(
\mathbf{A}_k,
\mathbf{Y}_k,
\mathbf{X}_k,
\mathbf{z}_{G},
\mathbf{O}_k^{\mathrm{que}}
\right),
\label{eqStructureFeature}
\end{equation}
where $\mathbf{O}_k^{\mathrm{que}}$ is the UAV queue state matrix. The feature $\boldsymbol{\xi}_k$ summarizes topology density, task load distribution, computation splitting pattern, link quality statistics and queue state statistics.

To encode the assignment relation, the task embeddings assigned to UAV $i$ are aggregated as
\begin{equation}
\bar{\mathbf{z}}_{i}^{T}
=
\frac{
\sum_{m\in\mathcal{M}_k^{\mathrm{rem}}}
y_{im}\mathbf{z}_m^{T}
}
{
\sum_{m\in\mathcal{M}_k^{\mathrm{rem}}}
y_{im}
+\epsilon
},
\quad
\forall i\in\mathcal{U}.
\label{eqPredictTaskToUav}
\end{equation}
The UAV representation used for prediction is updated by
\begin{equation}
\widehat{\mathbf{z}}_{i}^{U}
=
\mathbf{z}_{i}^{U}
+
\phi_{PU}
\left(
\left[
\left(\mathbf{z}_{i}^{U}\right)^{\top},
\left(\bar{\mathbf{z}}_{i}^{T}\right)^{\top}
\right]^{\top}
\right),
\quad
\forall i\in\mathcal{U},
\label{eqPredictUavUpdate}
\end{equation}
where $\phi_{PU}(\cdot)$ is a trainable fusion function. For task $m$, the embedding of its assigned sensing UAV is obtained as
\begin{equation}
\bar{\mathbf{z}}_{m}^{U}
=
\sum_{i\in\mathcal{U}}
y_{im}
\widehat{\mathbf{z}}_{i}^{U},
\quad
\forall m\in\mathcal{M}_k^{\mathrm{rem}}.
\label{eqPredictUavToTask}
\end{equation}
Then the task representation for prediction is updated by
\begin{equation}
\widehat{\mathbf{z}}_{m}^{T}
=
\mathbf{z}_{m}^{T}
+
\phi_{PT}
\left(
\left[
\left(\mathbf{z}_{m}^{T}\right)^{\top},
\left(\bar{\mathbf{z}}_{m}^{U}\right)^{\top}
\right]^{\top}
\right),
\quad
\forall m\in\mathcal{M}_k^{\mathrm{rem}},
\label{eqPredictTaskUpdate}
\end{equation}
where $\phi_{PT}(\cdot)$ is a trainable fusion function.

The predicted system completion time and task completion time are given by
\begin{equation}
\widehat T_k^{\mathrm{total}}
=
\left[
\phi_G
\left(
\left[
\mathbf{z}_{G}^{\top},
\boldsymbol{\xi}_k^{\top}
\right]^{\top}
\right)
\right]^+,
\label{eqPredictGlobalTime}
\end{equation}
\begin{equation}
\begin{aligned}
\widehat F_{m,k}
&=
\left[
\phi_M
\left(
\left[
\begin{array}{c}
\widehat{\mathbf{z}}_{m}^{T}\\
\bar{\mathbf{z}}_{m}^{U}\\
\mathbf{z}_{G}\\
\boldsymbol{\xi}_k
\end{array}
\right]
\right)
\right]^+,
\\
&\quad
m\in\mathcal{M}_k^{\mathrm{rem}},
\end{aligned}
\label{eqPredictTaskTime}
\end{equation}
where $\phi_G(\cdot)$ and $\phi_M(\cdot)$ are trainable prediction functions, and $[x]^+=\max\{x,0\}$ ensures nonnegative time prediction.

Before being used for decision selection, the prediction head is pretrained with supervised labels generated by task execution simulation. Let $T_k^{\mathrm{total}}$ and $F_{m,k}$ denote the simulated system completion time and task completion time under a sampled decision. The prediction loss is defined as the normalized squared error
\begin{equation}
\begin{aligned}
\mathcal{L}_{\mathrm{pred}}
&=
\left(
\frac{
\widehat T_k^{\mathrm{total}}-T_k^{\mathrm{total}}
}
{
|T_k^{\mathrm{total}}|+\epsilon
}
\right)^2
\\
&\quad+
\frac{1}{|\mathcal{M}_k^{\mathrm{rem}}|}
\sum_{m\in\mathcal{M}_k^{\mathrm{rem}}}
\left(
\frac{
\widehat F_{m,k}-F_{m,k}
}
{
|F_{m,k}|+\epsilon
}
\right)^2 .
\end{aligned}
\label{eqPredictionLoss}
\end{equation}

After pretraining, the prediction head is used to assist policy decision making. At each decision step, the policy network continuously samples $K$ candidate decisions\footnote{In this paper, $K=4$.}, denoted as $\{\mathcal{D}_k^{(1)},\ldots,\mathcal{D}_k^{(K)}\}$. Each candidate decision is fed into the prediction head to estimate the corresponding system completion time. The final decision is selected as
\begin{equation}
\mathcal{D}_k^{\star}
=
\arg\min_{\mathcal{D}_k^{(q)},\ q=1,\ldots,K}
\widehat T_k^{\mathrm{total}}
\left(
\mathcal{D}_k^{(q)}
\right).
\label{eqPredictionSelection}
\end{equation}
The selected decision $\mathcal{D}_k^{\star}$ is then executed by the environment. Therefore, the prediction module does not directly generate task allocation, topology or splitting decisions. Instead, it serves as a pretrained evaluator that selects the most promising candidate decision according to the predicted completion time.

\subsection{Reward Design}
The reward signal is designed according to the closed loop execution result induced by $\mathbf{Y}$, $\mathbf{A}$, $\mathbf{X}$, $\mathbf{P}$ and $\mathbf{B}$. It only considers three terms, namely the number of successfully completed tasks, the system completion time and the connectivity penalty.

Let $N_{\mathrm{succ}}$ denote the number of tasks completed before their deadlines. It is defined as
\begin{equation}
N_{\mathrm{succ}}
=
\sum_{m\in\mathcal{M}}\chi_m .
\label{eqRewardSuccessNumber}
\end{equation}
The system completion time is determined by the last finished task and is written as
\begin{equation}
T^{\mathrm{total}}
=
\max_{m\in\mathcal{M}}F_m .
\label{eqRewardTotalTime}
\end{equation}
The connectivity penalty is defined according to the algebraic connectivity of the generated communication graph as
\begin{equation}
P_{\mathrm{conn}}
=
\mathbf{1}
\left(
\lambda_2(\mathbf{L})\leq 0
\right),
\label{eqConnectivityPenalty}
\end{equation}
where $\mathbf{L}$ is the Laplacian matrix of the generated communication graph and $\lambda_2(\mathbf{L})$ denotes the algebraic connectivity.

The terminal reward of one episode is defined as
\begin{equation}
r
=
\omega_1 N_{\mathrm{succ}}
-
\omega_2 T^{\mathrm{total}}
-
\omega_3 P_{\mathrm{conn}},
\label{eqTerminalReward}
\end{equation}
where $\omega_1$, $\omega_2$ and $\omega_3$ are nonnegative coefficients. The first term encourages the policy to complete more tasks before their deadlines, the second term penalizes long system completion time, and the third term penalizes disconnected communication topology.

For policy learning, the value function is denoted as $V_{\psi}(\mathcal{G})$, where $\psi$ is the value network parameter. Let $\mathcal{D}_b=\{\mathbf{Y}_b,\mathbf{A}_b,\mathbf{X}_b\}$ denote the sampled decision of sample $b$. The advantage estimate is given by
\begin{equation}
\widehat A_b
=
r_b
-
V_{\psi}
\left(
\mathcal{G}_b
\right).
\label{eqRewardAdvantage}
\end{equation}
To constrain the policy update, PPO uses the probability ratio between the current policy and the old policy,
\begin{equation}
\rho_b(\theta)
=
\frac{
\pi_{\theta}
\left(
\mathcal{D}_b\mid\mathcal{G}_b
\right)
}
{
\pi_{\theta_{\mathrm{old}}}
\left(
\mathcal{D}_b\mid\mathcal{G}_b
\right)
}.
\label{eqPpoRatio}
\end{equation}
The clipped surrogate objective is written as
\begin{equation}
\begin{aligned}
\mathcal{J}_{\mathrm{clip}}(\theta)
&=
\frac{1}{B}
\sum_{b=1}^{B}
\min
\left(
\rho_b(\theta)\widehat A_b,
\,\right.
\\
&\quad\left.
\mathrm{clip}
\left(
\rho_b(\theta),
1-\epsilon_{\mathrm{ppo}},
1+\epsilon_{\mathrm{ppo}}
\right)
\widehat A_b
\right),
\end{aligned}
\label{eqPpoClipObjective}
\end{equation}
where $B$ is the batch size and $\epsilon_{\mathrm{ppo}}$ is the clipping threshold. The corresponding policy loss is defined as
\begin{equation}
\mathcal{L}_{\mathrm{policy}}
=
-
\mathcal{J}_{\mathrm{clip}}(\theta).
\label{eqPolicyLoss}
\end{equation}

The value loss is defined as
\begin{equation}
\mathcal{L}_{\mathrm{value}}
=
\frac{1}{B}
\sum_{b=1}^{B}
\left(
V_{\psi}
\left(
\mathcal{G}_b
\right)
-
r_b
\right)^2 .
\label{eqValueLoss}
\end{equation}
The total training loss is then written as
\begin{equation}
\mathcal{L}
=
\mathcal{L}_{\mathrm{policy}}
+
c_v\mathcal{L}_{\mathrm{value}}
+
c_p\mathcal{L}_{\mathrm{pred}}
-
\beta
\mathcal{H}
\left(
\pi_{\theta}
\right),
\label{eqTotalLoss}
\end{equation}
where $c_v$ is the value loss coefficient, $c_p$ is the prediction loss coefficient, $\beta$ is the entropy coefficient and $\mathcal{H}(\pi_{\theta})$ denotes the policy entropy. The prediction loss $\mathcal{L}_{\mathrm{pred}}$ is used to train the completion time prediction module and is not added to the environment reward.
\begin{figure*}[htbp]
\centering
\includegraphics[width=0.9\textwidth]{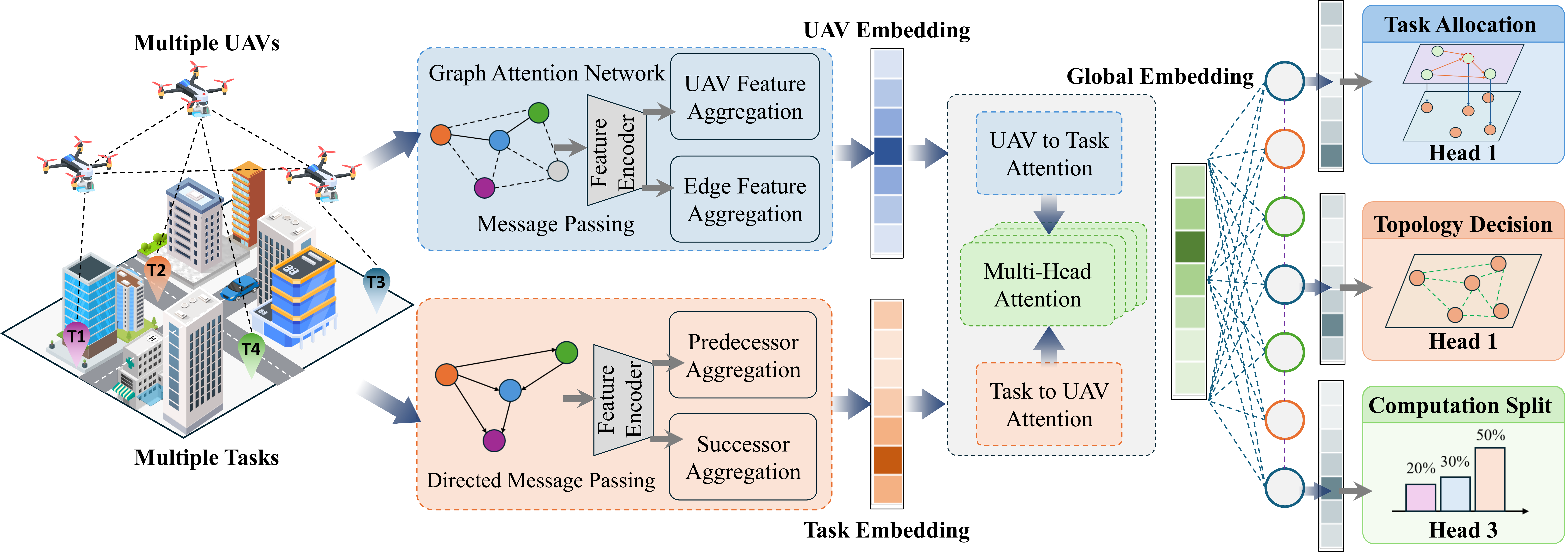}
\vspace{-8pt}
\caption{Overall architecture of the proposed dual heterogeneous graph learning framework. UAV and task graphs are encoded by graph attention and directed message passing, cross graph attention produces global embeddings, and three decision heads generate task allocation, topology decision and computation splitting.}
\label{figFramework}
\vspace{-12pt}

\end{figure*}

\section{Experimental Results}

\subsection{Numerical Results}

\subsubsection{Experimental Setup}
\paragraph{Parameters Setup}

The experiments are conducted in a simulated UAV sensing and computing scenario. The network consists of five UAVs deployed in a two dimensional area with multiple sensing tasks. The number of tasks varies from one to six during training and evaluation. The UAVs are characterized by mobility, computing, sensing and communication related attributes, while each task is described by its location, data size, computing workload and deadline. The main simulation parameters are summarized in Table~\ref{tab_simulation_parameters}.

\begin{table}[!t]
\centering
\caption{Simulation Parameters}
\label{tab_simulation_parameters}
\footnotesize
\setlength{\tabcolsep}{2.5pt}
\begin{tabular}{l c l c}
\hline
Parameter & Setting & Parameter & Setting \\
\hline
$N$ & 5 & $M$ & 1--6 \\
Area & $100~\mathrm{m}\times100~\mathrm{m}$ & $v_i^{\max}$ & $7.0$--$14.0~\mathrm{m/s}$ \\
$f_i$ & $6.0$--$14.0~\mathrm{GHz}$ & $B_i^{\max}$ & $18.0$--$42.0~\mathrm{MHz}$ \\
$P_i^{\max}$ & $0.65$--$1.65~\mathrm{W}$ & $L_m$ & $5.0$--$16.0~\mathrm{Mbit}$ \\
$c_m$ & $3.0$--$10.0~\mathrm{Gcycles/Mbit}$ & $\tau_m^{\max}$ & $24.0$--$58.0~\mathrm{s}$ \\
$\rho_i^{\mathrm{sen}}$ & $8.0~\mathrm{Mbit/s}$ & $R^{\min}$ & $0.10~\mathrm{Mbit/s}$ \\
$N_0$ & $0.018~\mathrm{W}$ & & \\
\hline
\end{tabular}
\vspace{-8pt}
\end{table}

The task set is modeled as a directed acyclic graph to describe precedence relationships among sensing tasks. The task graph complexity varies from 0.05 to 0.95 to evaluate the robustness of different methods under different dependency densities. The UAV communication candidate graph is generated according to the communication range and nearest neighbor augmentation, which preserves the influence of distance and link quality while keeping the graph connected in most random scenarios.

The proposed policy jointly determines the task allocation matrix, communication topology matrix and computing splitting ratios. UAVs assigned to sensing tasks collect data from target regions, while unassigned UAVs serve as computing UAVs to process the collected data. The policy network adopts dual graph message passing to encode the UAV communication graph and the task dependency graph, and then uses cross graph interaction to match UAV capabilities with task requirements.

The training process contains three stages. The completion time prediction module is first pretrained. The assignment reinforcement learning stage then optimizes the task allocation policy. Finally, the joint reinforcement learning stage optimizes task allocation, communication topology, computing splitting, value estimation and completion time prediction together. During evaluation, fixed testing seeds are used to compare all methods under the same scenario distribution.

The evaluation metrics include the number of completed tasks, system makespan, average task completion time, communication resource utilization, computing queue length, computing splitting ratio and reconfiguration count. The baselines include nearest distance assignment, resource priority assignment, random assignment and several ablation variants, including no splitting, mean splitting, no reconfiguration, graph neural network interaction and multilayer perceptron based decision making.

\paragraph{Benchmark Setup}

To evaluate the performance of the proposed method, the following benchmarks are introduced.

\begin{enumerate}

\item \textbf{No Split} uses the task allocation and communication topology generated by the proposed method, but task data are not allowed to be split and processed by other UAVs.

\item \textbf{Mean Split} uses the task allocation and communication topology generated by the proposed method, while task data are uniformly split among all feasible computing nodes.

\item \textbf{Graph Neural Network (GNN) PPO} replaces the attention based cross graph interaction between the communication graph and the task graph in the proposed method with equal weight message passing and feature aggregation.

\item \textbf{Multilayer Perceptron (MLP) PPO} is still trained by PPO and retains the same output forms for task allocation, communication topology, and computation splitting. However, message passing over the communication graph and the task graph is removed, and feature extraction and decision making are mainly performed through fully connected layers based on node features.

\item \textbf{Greedy} does not rely on neural network training. Instead, tasks are assigned sequentially according to their order. For each task to be assigned, all UAVs are enumerated, and the UAV with the minimum candidate completion time is selected as the executor. The communication topology is constructed according to inter UAV distance and resource scores, while computation splitting follows a uniform rule.

\end{enumerate}

\subsubsection{Results and Discussion}

\paragraph{Reward Convergence Analysis} 

\begin{figure}[!t]
\centering
\subfloat[Reward under different message passing rounds]{\includegraphics[width=0.8\columnwidth]{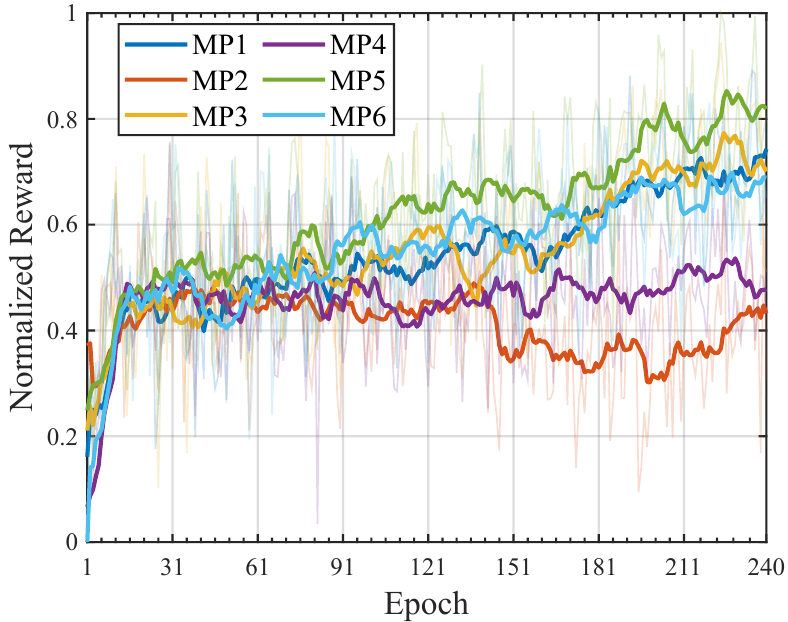}
\label{figRewardMp}}\\ \vspace{-6pt}
\subfloat[Reward comparison with baselines]{\includegraphics[width=0.8\columnwidth]{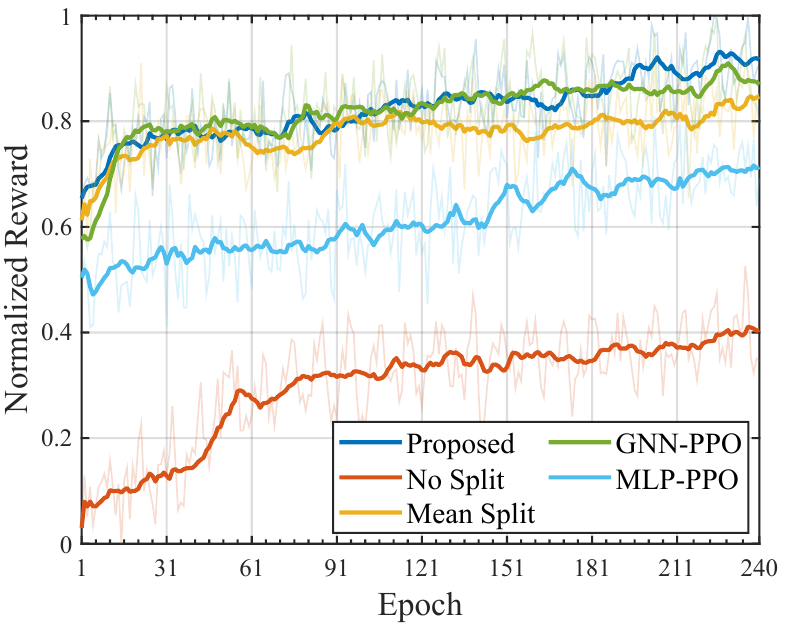}
\label{figRewardBaseline}}
\caption{Reward convergence under different architectural choices.}
\label{figRewardConvergence}
\end{figure}

Fig. \ref{figRewardConvergence}\subref{figRewardMp} illustrates the reward convergence under different numbers of message passing rounds. As the training process progresses, one and two message passing rounds aggregate information only from neighboring nodes. As a result, the policy has limited awareness of long range task dependencies and indirect communication relationships, leading to slower reward improvement and more pronounced fluctuations. Three and four message passing rounds expand the structural receptive field, enabling the network to better identify executable tasks and available computing nodes. Five message passing rounds achieve higher and more stable rewards, indicating that task dependencies in the task graph together with link states in the UAV graph are effectively propagated to the decision heads. Increasing the number of message passing rounds to six does not provide further performance improvement because excessive information propagation causes node representations to become overly similar, thereby reducing the discriminative capability of the learned features.

Fig. \ref{figRewardConvergence}\subref{figRewardBaseline} compares the reward evolution of the proposed method with different baseline methods. No Split consistently achieves the lowest rewards because relying solely on local computation overloads the queues of some UAVs, which increases waiting time and reduces the overall task completion reward. Mean Split alleviates computation congestion to a certain extent. However, uniform task partitioning does not account for link quality or node queue status, resulting in only limited reward improvement. GNN PPO performs worse than the proposed method, demonstrating the direct contribution of GAT to node feature aggregation. MLP PPO achieves substantially lower rewards, indicating that without message passing and feature aggregation, node representations cannot be effectively extracted, making it more difficult for the policy network to produce high quality decisions.

\paragraph{Task Completion Performance}

\begin{figure*}[!t]
\centering
\subfloat[Makespan versus task number]{\includegraphics[width=0.5\columnwidth]{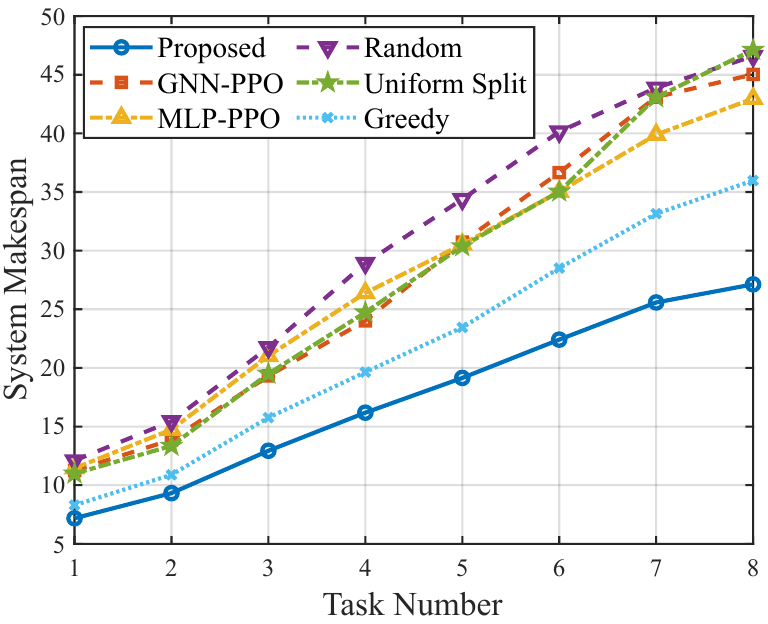}
\label{figTimeTaskNumber}}
\subfloat[Successful tasks versus task number]{\includegraphics[width=0.5\columnwidth]{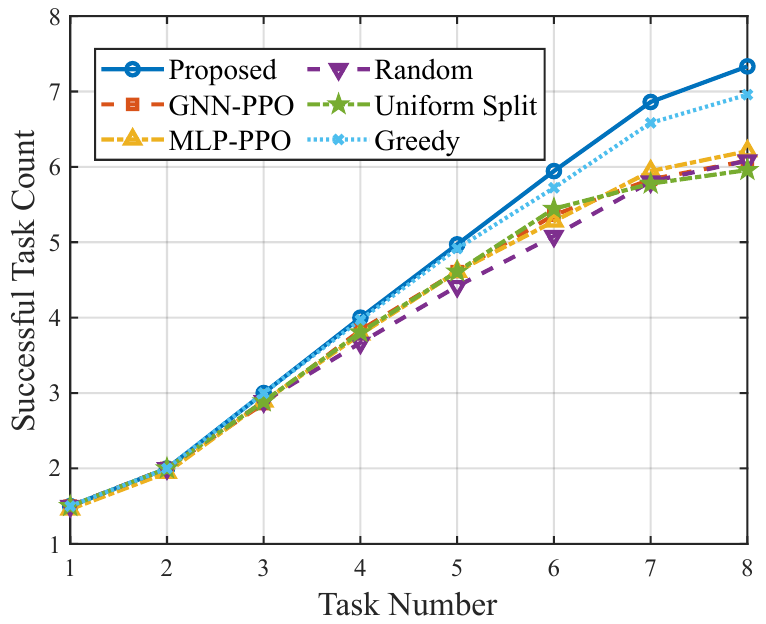}
\label{figSuccessTaskNumber}}
\subfloat[Makespan versus graph complexity]{\includegraphics[width=0.5\columnwidth]{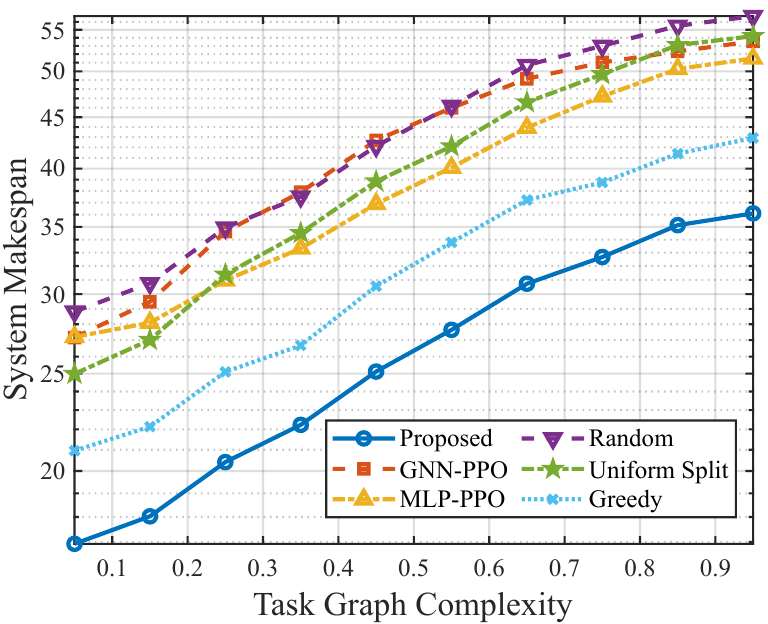}
\label{figTimeGraphComplexity}}
\subfloat[Successful tasks versus graph complexity]{\includegraphics[width=0.5\columnwidth]{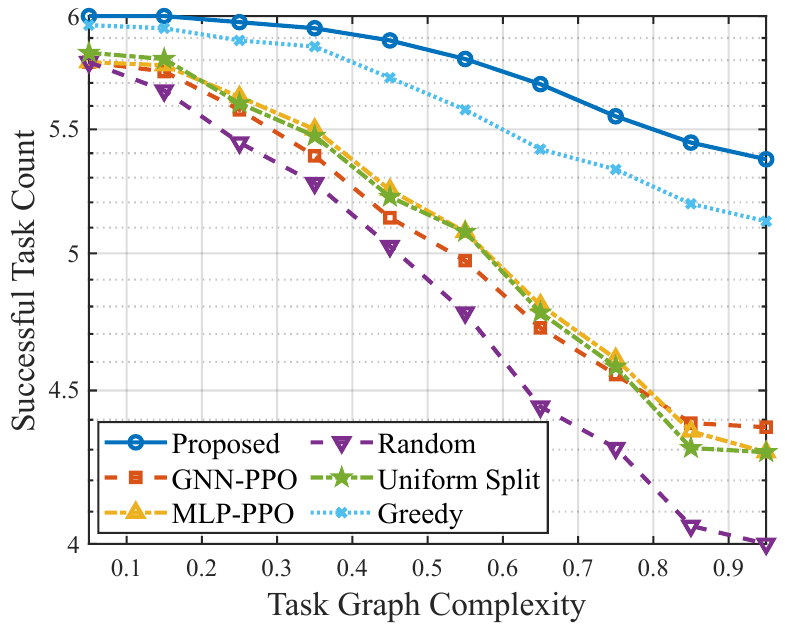}
\label{figSuccessGraphComplexity}}\vspace{-5pt}
\caption{Task completion performance under changing task number and task graph complexity.}
\label{figCompletionPerformance}
\vspace{-5pt}
\end{figure*}

Fig. \ref{figCompletionPerformance}\subref{figTimeTaskNumber} presents the system makespan under different numbers of tasks. As the number of tasks increases, the makespan of all methods rises because more tasks introduce longer sensing workflows, higher data transmission demand, and heavier computation queues. The proposed method consistently achieves the lowest makespan, demonstrating that the joint optimization of task allocation, topology selection, and computation partitioning effectively distributes task data to more suitable UAV nodes. This reduces queueing pressure on individual UAVs while preventing low quality communication links from carrying excessive transmission traffic.

Fig. \ref{figCompletionPerformance}\subref{figSuccessTaskNumber} shows the number of successfully completed tasks under different task scales. As the task scale increases, the baseline methods become more susceptible to computation congestion and communication delay, limiting the growth in the number of completed tasks. The proposed method consistently completes more tasks under large task scales, indicating that it maintains a more effective match between task requirements and UAV resource states. Since both task allocation and computation partitioning adapt to the network load, critical tasks in the dependency chain are more likely to be completed before their deadlines.

Fig. \ref{figCompletionPerformance}\subref{figTimeGraphComplexity} illustrates the impact of task graph complexity on the makespan. Higher graph complexity introduces more dependency edges, requiring additional predecessor tasks to finish before subsequent tasks can start. This also increases the amount of data exchanged among UAVs. The makespan of the baseline methods increases more rapidly, indicating that they struggle to jointly coordinate task execution order and network resource allocation. In contrast, the proposed method exhibits a slower increase, suggesting that the dual graph representation enables the policy to identify critical dependency paths and establish communication links that better support subsequent computation.

Fig. \ref{figCompletionPerformance}\subref{figSuccessGraphComplexity} presents the number of successfully completed tasks under different task graph complexities. As the graph complexity increases, local delays propagate more easily along dependency chains, resulting in fewer completed tasks. The proposed method experiences a smaller performance degradation, indicating that it maintains stronger task execution capability under complex dependency structures. This result further demonstrates that the joint matching between the communication graph and the task graph effectively reduces task failures caused by insufficient communication links and excessive computation queueing.

\paragraph{Prediction Accuracy Analysis}

\begin{figure}[htbp]
\centering
\includegraphics[width=0.8\columnwidth]{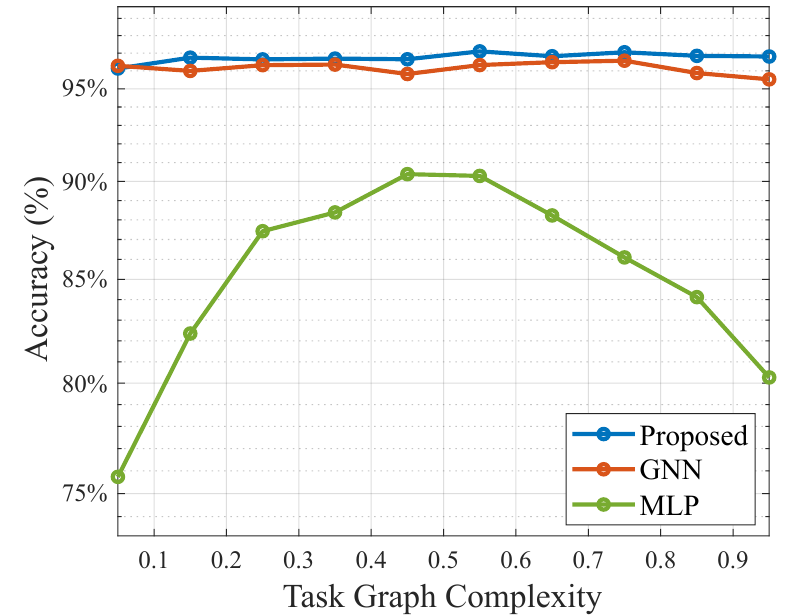}
\vspace{-6pt}
\caption{Completion time prediction accuracy under varying task graph complexity. The proposed predictor and the GNN based predictor remain stable, while the MLP based predictor is more sensitive to dependency density.}
\label{figPredictionAccuracyComplexity}
\end{figure}

Fig. \ref{figPredictionAccuracyComplexity} illustrates the completion time prediction accuracy under different task graph complexities. The proposed method and GNN maintain high prediction accuracy across the entire complexity range, indicating that jointly encoding the task graph and the UAV graph effectively captures the structural characteristics of both the communication graph and the task graph. In contrast, the accuracy of MLP decreases noticeably in both low and high complexity scenarios because vector based representations cannot effectively distinguish critical dependency paths or local network bottlenecks. These results indicate that the prediction module must explicitly exploit graph structure to provide reliable risk assessment for subsequent network reconfiguration. Without structural inductive bias for the task graph and the UAV communication graph, the MLP prediction head tends to learn an average completion time pattern across different graph complexities. Consequently, it overestimates the completion time of highly parallel tasks in low complexity scenarios, while underestimating the delay introduced by long dependency chains and accumulated computation queues in high complexity scenarios.

\vspace{-6pt}
\paragraph{Impact of Message Passing Rounds}

\begin{figure*}[!t]
\centering
\subfloat[Makespan]{\includegraphics[width=0.3\textwidth]{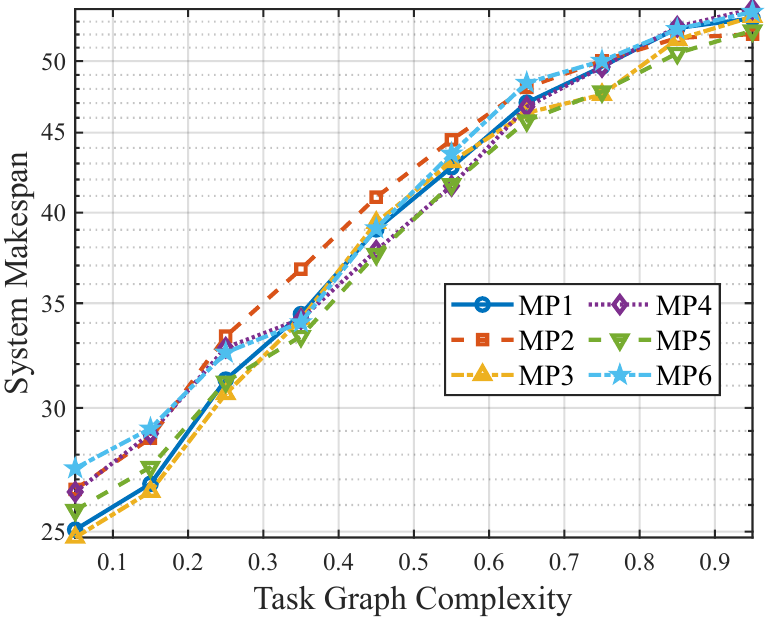}
\label{figMpMakespan}}
\subfloat[Successful tasks]{\includegraphics[width=0.3\textwidth]{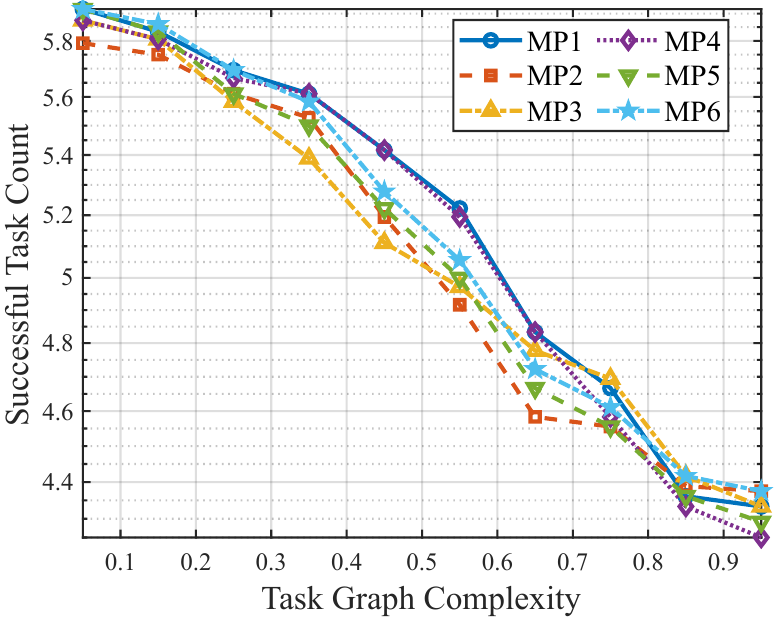}
\label{figMpSuccess}}
\subfloat[Prediction accuracy]{\includegraphics[width=0.3\textwidth]{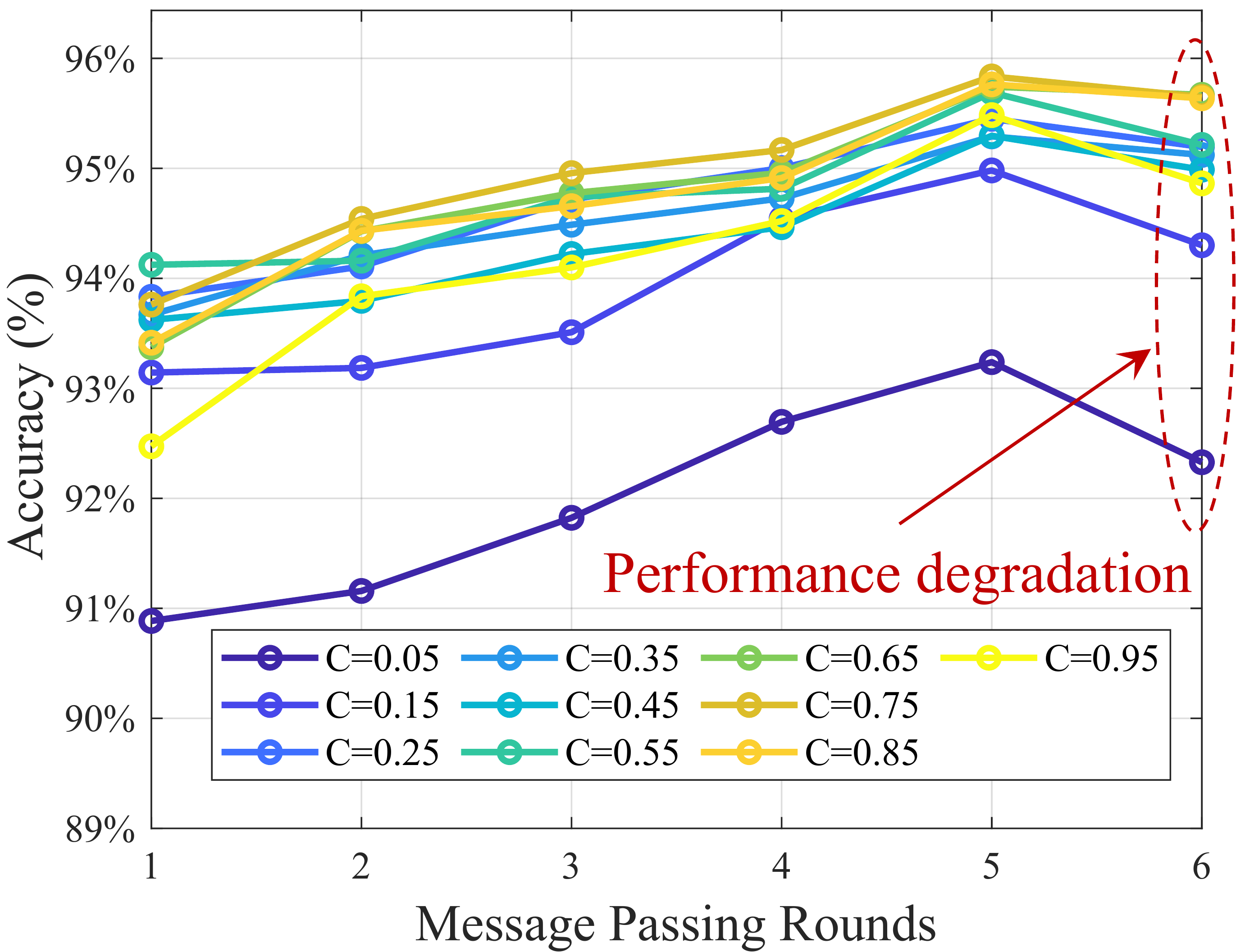}
\label{figMpAccuracy}}\vspace{-6pt}
\caption{Impact of message passing rounds on task execution and prediction performance.}
\label{figMessagePassingRounds}
\vspace{-6pt}
\end{figure*}

Fig. \ref{figMessagePassingRounds}\subref{figMpMakespan} and Fig. \ref{figMessagePassingRounds}\subref{figMpSuccess} illustrate the impact of the number of message passing rounds on the makespan and the number of successfully completed tasks. A small number of message passing rounds is sufficient for low complexity scenarios but fails to capture dependency information propagated through long multi hop paths in complex task graphs. Five message passing rounds generally achieve the lowest makespan and the highest number of completed tasks, indicating that an appropriate propagation depth effectively covers critical task relationships and improves the joint perception of task dependencies and communication connectivity.

Fig. \ref{figMessagePassingRounds}\subref{figMpAccuracy} presents the prediction accuracy under different numbers of message passing rounds. As the number of message passing rounds increases from one to five, the prediction accuracy consistently improves, indicating that more comprehensive structural information propagation enables the prediction module to better identify waiting time introduced by task dependencies as well as communication bottlenecks. However, the prediction performance deteriorates when the number of message passing rounds increases to six. This result suggests that excessive information propagation causes node representations to become overly similar, thereby weakening the feature extraction capability of the network.

\subsection{Application Case}

\subsubsection{Experimental Setup}
To verify the execution feasibility of the proposed method, an AirSim based validation platform is constructed. Five multirotor UAVs are configured in the experiment, and their takeoff, trajectory tracking, image acquisition, and task execution are uniformly controlled through application programming interfaces (APIs). The validation process starts from natural language task requirements. The Qwen-VL-plus model is first invoked to transform the user input into a structured task graph, where each node represents a sensing task and each directed edge denotes the precedence dependency between tasks. Then, the task graph adapter converts the JavaScript Object Notation (JSON) based task description into task node features and dependency edge representations required by the trained model, which are further fed into the proposed dual graph decision model.

\begin{figure*}[t!]
\centering
\includegraphics[width=0.8\textwidth]{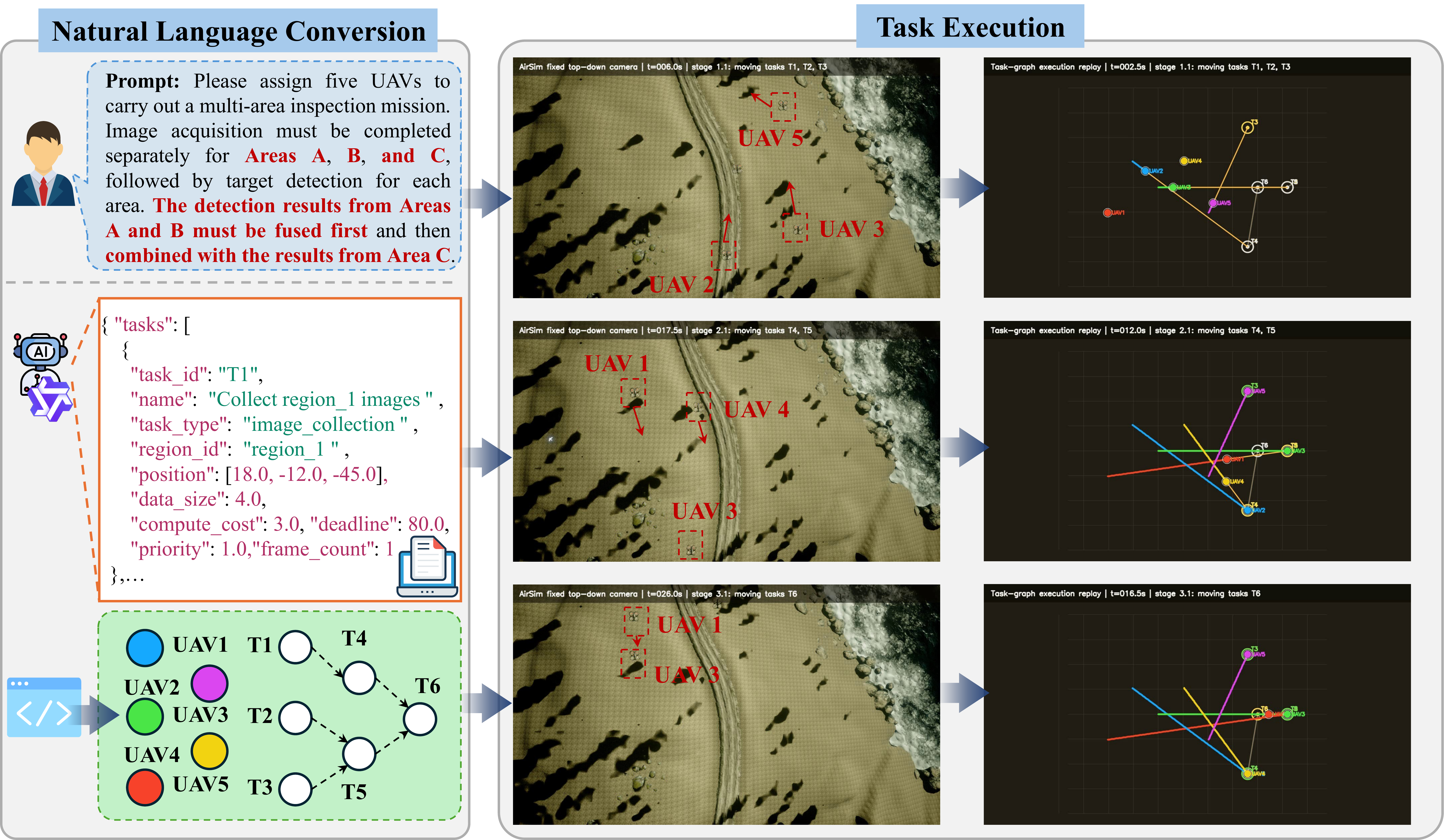}
\vspace{-8pt}
\caption{Application case of task graph guided multi UAV collaborative execution from natural language instructions.}
\label{Sim2Real_Res_BG}
\vspace{-2pt}
\end{figure*}

During execution, the five UAVs first take off from a predefined initial formation and climb to a unified safe altitude. They then fly to the corresponding task regions according to the task allocation results generated by the model. After reaching the target regions, the UAVs collect red, green, and blue (RGB) image sequences through the image interface, and You Only Look Once (YOLO) inference is adopted to perform object detection. For the image sequence collected for each task, the frame indices are assigned to different UAVs according to the computation splitting ratios generated by the model, as shown in Fig. \ref{Sim2Real_Res_BG}.

\subsubsection{Results and Discussion}
\begin{figure}[t!]
\centering
\includegraphics[width=0.8\columnwidth]{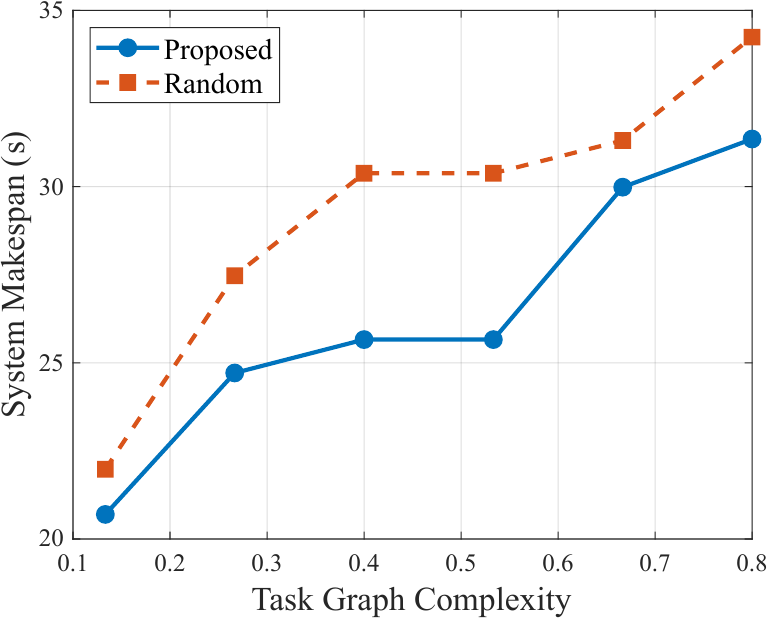}
\vspace{-6pt}
\caption{Makespan versus graph complexity in AirSim simulation.}
\label{Sim2Real_Res}
\vspace{-8pt}
\end{figure}

As shown in Fig. \ref{Sim2Real_Res}, the proposed method completes the full closed loop process from natural language task generation to task graph construction and multi UAV collaborative execution in the simulation platform, while achieving a shorter task completion time. This indicates that joint task allocation, communication topology selection, and computation splitting effectively reduce the computation burden on individual UAVs and improve the execution efficiency of multi UAV sensing and computing tasks.

\section{Conclusion}

The paper proposes a dual heterogeneous graph learning based task allocation method for multi UAV systems with task dependencies. A task graph and a UAV graph are constructed to formulate task allocation as a structural matching problem between two graphs. A GAT together with a cross attention mechanism and PPO is employed to optimize task allocation, topology decision and computation data offloading jointly. Simulation results demonstrate that the proposed method achieves more completed tasks and shorter task completion time under different task scales and task complexity levels. Furthermore, an AirSim based UAV sensing and computing case integrated with an LLM is developed to automatically convert natural language task requirements into a task graph, further verifying the effectiveness and engineering potential of the proposed method.

\vspace{-12pt}
\bibliographystyle{IEEEtran}
\bibliography{IEEEfull}

\end{document}